\documentclass[a4paper]{cas-dc}
\usepackage[numbers,sort&compress]{natbib}
\usepackage{url}
\usepackage{graphicx}
\usepackage{float}
\usepackage{caption}
\usepackage{subcaption}%
\usepackage{tabularray}
\usepackage{enumitem}
\usepackage{algorithmic}
\usepackage{cleveref}
\usepackage[linesnumbered,ruled,vlined]{algorithm2e}
\usepackage{booktabs}
\usepackage{siunitx}
\usepackage{array}
\usepackage{multirow}
\usepackage{makecell}
\usepackage{amssymb}
\usepackage{adjustbox}

\usepackage[T1]{fontenc}
\usepackage{tabularx}
\usepackage{wasysym}%
\usepackage{xcolor}


\newcolumntype{C}[1]{>{\centering\arraybackslash}m{#1}}

\renewcommand{\tabularxcolumn}[1]{m{#1}}

\newcolumntype{Y}{>{\raggedright\arraybackslash}X}
\newcommand{\yes}{\CIRCLE}%
\newcommand{\partvis}{\LEFTcircle}%
\newcommand{\no}{\Circle}%
\newcommand{\notpresent}{\textemdash}%
\newcommand{\recoverable}{\ensuremath{\Diamond}}%

\let\printorcid\relax

\begin{document}
\let\WriteBookmarks\relax

\shorttitle{FLARE}

\shortauthors{Tong Guan et~al.}

\title [mode = title]{When Handshakes Go Dark: An Adaptive Semantic-Sequential Fusion Framework for Encrypted Traffic Classification}

\author[1]{Tong Guan}[style=chinese]

\author[1]{Yunchun Li}[style=chinese]

\author[1]{Minyuan Xiao}[style=chinese]

\author[2]{Mingyuan Xia}[style=chinese]

\author[1]{Yuchen Zhao}[style=chinese]

\author[1]{Wei Li}[style=chinese]\cormark[1]

\affiliation[1]{organization={School of Computer Science and Engineering, Beihang University},
    city={Beijing},
    country={China}}

\affiliation[2]{organization={School of Cyber Science and Technology, Beihang University},
    city={Beijing},
    country={China}}

\cortext[cor1]{Corresponding author}

\begin{abstract}
Encrypted traffic classification is essential for network security monitoring and traffic management. Self-supervised pre-training provides a promising approach to learning reusable representations from large-scale unlabeled traffic data. However, encrypted communication protocols such as TLS~1.3 and QUIC protect increasing portions of application and handshake data, while privacy-enhancing mechanisms such as Encrypted Client Hello (ECH) further conceal service-identifying metadata. Consequently, observable upper-layer bytes are increasingly dominated by ciphertext, weakening the correspondence between raw-byte patterns and application semantics. Raw-byte-based models therefore struggle to learn generalizable, semantically meaningful representations from strongly encrypted traffic. Moreover, the availability and discriminative value of observable information vary across protocols, protection profiles, and classification tasks, making it difficult to maintain effective
classification across diverse traffic scenarios. To address these challenges, we propose FLARE, a \textbf{F}ield-aware \textbf{L}earning and \textbf{A}daptive \textbf{R}epresentation Fusion framework for \textbf{E}ncrypted Traffic Classification. FLARE pre-trains a BERT-based packet encoder using Masked Field Prediction to reconstruct masked tokens from field-structured context. During downstream fine-tuning, the parameters of the pre-trained packet encoder remain frozen, and a packet-level gated fusion module adaptively combines protocol-semantic representations with directional packet-length representations, while a flow-level encoder models inter-packet dependencies for classification. We evaluate FLARE on eight datasets covering diverse encrypted-traffic classification tasks, including CipherSpectrum and VisQUIC datasets, which represent strongly encrypted traffic scenarios. FLARE achieves superior performance over baselines across the evaluated datasets. Notably, it outperforms the strongest pre-trained baseline by 10.92\% on the QUIC-based website fingerprinting task, highlighting the effectiveness of field-aware pre-training and adaptive semantic-sequential fusion under constrained observability.
\end{abstract}


\maketitle

\section{Introduction}
Traffic classification is a fundamental capability for network security monitoring and management, supporting application and service identification, quality-of-service assurance, and anomalous and malicious traffic detection~\cite{aouedi2025deep}. Meanwhile, encrypted communication has become the norm on the modern web, with HTTPS accounting for the vast majority of browser page loads and TLS~1.3 and QUIC constituting a substantial share of secure HTTP communication~\cite{google2024,cloudflare_radar_2026}. In a conventional full TLS~1.2 handshake, a passive observer can access plaintext handshake information such as the Server Name Indication (SNI), offered cipher suites, and the server certificate~\cite{rfc5246,rfc6066}. TLS~1.3 reduces the handshake information exposed to passive observers by encrypting the server certificate and the handshake messages following the ServerHello~\cite{rfc9846}. When ECH is successfully negotiated, the service-identifying SNI and other sensitive ClientHello parameters are further concealed~\cite{rfc9849}. QUIC further encrypts most transport-control information carried in frames and applies header protection to packet numbers and selected header bits, leaving only a limited wire image visible to passive observers~\cite{rfc9001,rfc9312}. Table~\ref{tab:field-visibility} summarizes how these protection mechanisms reduce and reshape the observability of application-layer content and service-identifying handshake information. This shift substantially constrains the information available for traffic classification and challenges methods that rely on plaintext content or high-level handshake features.

\begin{table*}[t]
\centering

\caption{
Observability of application-layer information, protocol-layer metadata, and behavioral metadata across representative protection profiles.
}
\label{tab:field-visibility}

\footnotesize
\setlength{\tabcolsep}{2.1pt}
\renewcommand{\arraystretch}{1.02}

\begin{tabularx}{\textwidth}{
    @{}
    C{2.50cm}
    Y
    C{1.15cm}
    C{1.18cm}
    C{1.18cm}
    C{1.35cm}
    C{1.15cm}
    @{}
}

\toprule

\makecell[c]{\bfseries Information\\\bfseries Category}
&
\makecell[c]{\bfseries Representative\\\bfseries Information}
&
\makecell[c]{\scriptsize\bfseries Plain\\
             \scriptsize\bfseries HTTP}
&
\makecell[c]{\scriptsize\bfseries TLS 1.2}
&
\makecell[c]{\scriptsize\bfseries TLS~1.3}
&
\makecell[c]{\scriptsize\bfseries TLS~1.3\\
             \scriptsize\bfseries + ECH}
&
\makecell[c]{\scriptsize\bfseries QUIC}
\\

\midrule

\multirow[c]{2}{2.05cm}[2.5ex]{
    \makecell[c]{Application-Layer\\Information}
}
&
\textbf{HTTP Request semantics and headers:}
method, target, authority, and other header fields
&
\yes & \no & \no & \no & \no
\\
\cmidrule(lr){2-7}
&
\textbf{Application payload}
&
\yes & \no & \no & \no & \no
\\

\midrule

\multirow[c]{7}{2.05cm}[-6.5ex]{
    \makecell[c]{Protocol-Semantic\\Metadata}
}
&
\textbf{Network-layer fields:}
version, header length, DSCP/ECN, total length,
fragmentation, TTL/hop limit, next protocol, and addresses
&
\yes & \yes & \yes & \yes & \yes
\\
\cmidrule(lr){2-7}
&
\textbf{TCP header fields:}
ports, header length, control flags,
sequence and acknowledgment numbers, window size, checksum, and options
&
\yes & \yes & \yes & \yes & \notpresent
\\
\cmidrule(lr){2-7}
&
\textbf{UDP header fields:}
ports, datagram length, and checksum
&
\notpresent & \notpresent & \notpresent & \notpresent & \yes
\\

\cmidrule(lr){2-7}

&
\textbf{TLS record-layer framing:}
record boundaries, outer record header,
and ciphertext length
&
\notpresent
&
\yes
&
\partvis\textsuperscript{1}
&
\partvis\textsuperscript{1}
&
\notpresent
\\
\cmidrule(lr){2-7}
&
\textbf{ClientHello fields:}
SNI, ALPN, supported versions, cipher suites, and extensions
&
\notpresent
&
\yes
&
\yes
&
\partvis\textsuperscript{2}
&
\recoverable\textsuperscript{3}
\\
\cmidrule(lr){2-7}
&
\textbf{Server certificate and authentication-related handshake messages}
&
\notpresent
&
\yes\textsuperscript{4}
&
\no
&
\no
&
\no
\\

\cmidrule(lr){2-7}

&
\textbf{QUIC wire-image fields:}
header form, version, and packet type in long headers, connection IDs, long-header length, and selected public bits
&
\notpresent
&
\notpresent
&
\notpresent
&
\notpresent
&
\partvis\textsuperscript{5}
\\

\midrule

\makecell[c]{Behavioral\\Metadata}
&
Packet length, direction, observed order and timing
&
\yes & \yes & \yes & \yes & \yes
\\

\bottomrule

\end{tabularx}

\vspace{1pt}

\noindent
\begin{minipage}{\textwidth}
\fontsize{7}{8}\selectfont
\raggedright
\setlength{\parindent}{0pt}
\setlength{\parskip}{0pt}

\textbf{Profiles:}
Plain HTTP denotes HTTP/1.1 over TCP;
the TLS columns denote HTTP over TLS/TCP;
and QUIC denotes HTTP/3 over QUIC without ECH.

\textbf{Symbols:}
\yes{} directly observable and parseable;
\partvis{} partially observable or limited to outer framing/metadata;
\recoverable{} recoverable from publicly derivable keys or protocol information;
\no{} cryptographically protected from a passive observer; and
\notpresent{} corresponding protocol element not present.

\textsuperscript{1}
TLS~1.3 exposes outer record framing and ciphertext length, while the inner content type and post-\texttt{ServerHello} messages are encrypted.

\textsuperscript{2}
ECH encrypts the true SNI and other sensitive ClientHello parameters; \texttt{ClientHelloOuter} remains visible and may expose public
parameters.

\textsuperscript{3}
QUIC Initial contents are recoverable using publicly derivable keys; later encryption levels remain protected.

\textsuperscript{4}
Assumes a full certificate-based TLS~1.2 handshake without session resumption; post-\texttt{ChangeCipherSpec} messages remain protected.

\textsuperscript{5}
QUIC long headers expose the version, packet type, connection IDs, and the \texttt{Length} field; short headers expose less, while packet numbers and selected header bits are protected.

\end{minipage}

\end{table*}

Early traffic classifiers typically combine manually engineered flow-level statistics with conventional machine-learning models. Representative methods such as AppScanner and BIND use features including packet and byte counts, flow duration, and inter-arrival times~\cite{taylor2016appscanner,al2016adaptive}. Although compact, these representations depend on domain expertise and handcrafted feature engineering.

Deep-learning methods subsequently reduced this dependence by learning representations directly from traffic data. DeepPacket, for example, learns from raw byte sequences~\cite{lotfollahi2020deep}. FS-Net captures communication behavior from packet-length and packet-direction sequences~\cite{liu2019fs}, and Deep Fingerprinting uses packet-direction sequences~\cite{sirinam2018deep}. These sequence features remain passively observable even when application-layer payloads are encrypted, and preserve useful discriminative information. However, these methods generally require substantial labeled traffic, limiting their adaptability to new applications, protocols, and network environments.

Pre-training offers a promising paradigm for reducing label dependency through self-supervised learning on large volumes of unlabeled traffic. Existing studies have explored representations at the byte, packet, burst, and flow levels. ET-BERT adopts masked prediction over byte tokens~\cite{lin2022bert}, Flow-MAE learns traffic representations through masked reconstruction~\cite{hang2023flow}, NetGPT introduces generative traffic pre-training~\cite{meng2023netgpt}, and TrafficFormer further models inter-packet sequential relationships~\cite{zhou2025trafficformer}.

Despite recent progress, traffic classification under strong encryption still faces two key challenges. First, modern protection mechanisms substantially reduce and reshape the observable features. As summarized in Table~\ref{tab:field-visibility}, application content and service-identifying handshake metadata become increasingly unavailable. Under such constrained observability, the correspondence between observable byte patterns and application semantics is substantially weakened, making it difficult to extract semantically meaningful and discriminative representations from ciphertext-dominated traffic bytes. Second, the availability and discriminative value of observable information vary across protocols, encryption profiles, and downstream classification tasks. Such variation arises even within a single connection, where handshake records expose parsable protocol fields but subsequent encrypted records do not. This variation makes it difficult to maintain effective classification across diverse traffic scenarios. These challenges motivate an approach that learns informative representations from the information that remains observable under encryption and integrates these representations adaptively at the packet level.

To address these limitations, we propose FLARE, a \textbf{F}ield-aware \textbf{L}earning and \textbf{A}daptive \textbf{R}epresentation fusion framework for \textbf{E}ncrypted traffic classification. At the packet level, FLARE parses packets into protocol fields, constructs field-aligned token sequences, and pre-trains a BERT-based encoder using a Masked Field Prediction (MFP) task. MFP masks selected tokens in the field-structured packet sequence and reconstructs them from the surrounding protocol context. Rather than treating all byte positions uniformly, the objective uses protocol-defined field organization to guide representation learning, encouraging the encoder to capture field-aware contextual dependencies. During downstream fine-tuning, the pre-trained encoder, referred to as PacketEncoder, is frozen; that is, its pre-trained parameters are not updated during task-specific fine-tuning and remain unchanged at inference. The frozen PacketEncoder produces protocol-semantic representations for individual packets. A gated fusion module combines each representation with its corresponding directional packet-length representation, and a flow-level encoder models the resulting sequence for task-specific classification. This separation allows reusable packet-level protocol knowledge to be adapted to different downstream tasks.

The main contributions are summarized as follows:
\begin{itemize}

    \item \textbf{Field-aware packet-level pre-training.} We construct packet-level inputs by parsing protocol fields across the network, transport, and encryption-protocol layers, including the TLS record and QUIC headers, while excluding identity-related fields and retaining only a short payload prefix. We pre-train a packet encoder with Masked Field Prediction, which aligns the prediction targets and their context with protocol fields, so that the encoder learns field values, inter-field dependencies, and packet structure that remain observable under encryption. The pre-trained encoder is then frozen and reused across downstream tasks.

       \item \textbf{Fine-grained adaptive semantic-sequential fusion.} Since the observable information varies across tasks and even between packets of the same connection, FLARE introduces a packet-level, feature-wise gating mechanism conditioned on both the protocol-semantic and directional packet-length representations of each packet. The gate dynamically recalibrates the contribution of each representation at the level of individual feature dimensions, allowing the fusion to adapt to the observability of each packet.

    \item \textbf{Extensive evaluation across diverse encryption scenarios.} We evaluate FLARE on eight datasets covering application classification, malicious traffic detection, website fingerprinting, and encrypted network service classification tasks. The evaluation includes CipherSpectrum~\cite{wickramasinghe2025sok} and VisQUIC~\cite{gahtan2024exploring}, which represent strongly encrypted traffic scenarios where the observability of application content and service-identifying handshake metadata is substantially restricted. FLARE achieves superior performance over a diverse set of baselines, with particularly pronounced gains under constrained observability.
\end{itemize}

The source code and pre-trained models of FLARE are publicly available at \url{https://github.com/sunshine12160110/FLARE}.

\section{Related Work}
\label{sec:relatedwork}
\subsection{Traditional Machine Learning-Based Traffic Classification}
As port-based and payload-inspection rules were rendered ineffective by encryption and port obfuscation, early studies relied on manually engineered flow statistics and traffic fingerprints combined with conventional machine-learning classifiers. AppScanner extracts statistical features from packet length and timing patterns for mobile application identification, while BIND constructs bidirectional flow representations for traffic classification~\cite{taylor2016appscanner,al2016adaptive}. CUMUL constructs traffic fingerprints from encrypted website traces and uses them for website fingerprinting~\cite{panchenko2016website}.

These approaches demonstrate that useful traffic characteristics can be derived without decrypting application content. However, their effectiveness depends heavily on domain knowledge and handcrafted feature engineering. Moreover, the distributions of engineered statistics and fingerprints may vary with network conditions, capture environments, application behavior, and traffic composition, making it difficult to design a fixed handcrafted feature set that remains effective across diverse traffic scenarios.

\subsection{Deep Learning-Based Traffic Classification}
Deep learning~\cite{liu2019fs, wang2017end, wang2017malware, lotfollahi2020deep, fauvel2023lightweight, shen2021accurate, sirinam2018deep} substantially reduced the reliance on manual feature engineering by learning traffic representations directly from traffic data. One line of research operates on raw byte sequences. DeepPacket employs convolutional and autoencoder-based architectures to learn end-to-end representations from packet bytes~\cite{lotfollahi2020deep}. Such methods can discover discriminative byte-level patterns automatically. However, as modern protocols encrypt increasing portions of upper-layer content, the correspondence between observable byte patterns and application semantics becomes progressively weaker.

Another line of research models communication behavior that remains observable without payload decryption. FS-Net learns flow representations from packet-length and packet-direction sequences, while Deep Fingerprinting models direction-aware packet sequences for website fingerprinting~\cite{liu2019fs,sirinam2018deep}. LEXNet further develops lightweight sequence modeling for encrypted traffic classification~\cite{fauvel2023lightweight}. Graph-based approaches such as GraphDApp organize packet interactions into structured representations to capture relational characteristics within traffic sessions~\cite{shen2021accurate}. These studies demonstrate the effectiveness of raw-byte, behavioral, and relational representations for encrypted traffic analysis. Packet-length, direction, and timing sequences remain passively observable under encryption, although their distributions may be affected by network conditions.

Nevertheless, most of these models are trained for specific tasks and require substantial labeled traffic, limiting their adaptation to emerging applications, protocols, and deployment environments.

\subsection{Pre-training-Based Traffic Classification}

Self-supervised pre-training enables traffic models to learn reusable representations from large volumes of unlabeled data before adapting to downstream tasks. Existing approaches have explored diverse backbone architectures, traffic representations, and pre-training objectives.

Transformer-based methods~\cite{lin2022bert,he2020pert, lin2021pean,lin2024cetp, peng2024ptu, zhou2025trafficformer, chen2025miett, yang2025mm4flow} constitute a major line of traffic pre-training research. ET-BERT tokenizes bidirectional traffic bursts and employs traffic-specific masked and contextual prediction tasks to learn contextualized datagram representations~\cite{lin2022bert}. TrafficFormer further incorporates fine-grained traffic objectives and inter-packet order and direction relationships into Transformer-based pre-training~\cite{zhou2025trafficformer}. MIETT treats packets as separate instances within a flow and uses two-level attention to capture both token-level and packet-level relationships. Its Packet Relative Position Prediction and Flow Contrastive Learning objectives further model temporal and flow-level structures~\cite{chen2025miett}.

Another major line of methods formulates traffic pre-training as masked autoencoding. YaTC converts raw flows into formatted matrices containing multi-level flow information and employs packet-level and flow-level attention within a masked autoencoder framework~\cite{zhao2023yet}. Flow-MAE represents traffic bursts as patch sequences and reconstructs masked regions to learn compact representations for malicious traffic classification~\cite{hang2023flow}. These approaches demonstrate that matrix-based and patch-based representations can preserve hierarchical traffic information while enabling efficient masked reconstruction.

Generative and efficient architectures have also been introduced into traffic pre-training. NetGPT formulates traffic understanding and generation within a unified generative pre-training framework using multi-pattern traffic representations~\cite{meng2023netgpt}. NetMamba and PaT replace quadratic-complexity self-attention with linear-time state-space modeling and develop traffic representations for efficient long-sequence processing~\cite{wang2024netmamba, song2026pat}. These studies extend traffic pre-training beyond encoder-only Transformers and explore broader trade-offs among representation capacity, efficiency, and downstream adaptability.

Recent studies have further extended traffic pre-training to multi-granularity and multimodal representations. MLETC jointly learns representations at the byte, field, and packet levels, with field-level learning applied to the IP and TCP/UDP headers~\cite{park2025multi}. MM4flow separately pre-trains representations of the payload bytes and the packet-length sequence and integrates them through cross-modal fusion during downstream fine-tuning~\cite{yang2025mm4flow}. These studies integrate traffic information at multiple granularities or from multiple modalities to enrich traffic representations.

However, modern encryption protocols continue to restrict and reshape the information available to passive traffic classifiers. TLS 1.3 and QUIC encrypt or conceal an increasing proportion of application-layer content and handshake metadata, while protocol header fields and packet length and direction remain largely observable. The availability and discriminative value of the remaining observable features vary across traffic scenarios and downstream tasks. Learning informative and reusable traffic representations under such constrained and variable observability therefore remains an open challenge.

\begin{figure*}
\centering
\includegraphics[width=\textwidth]{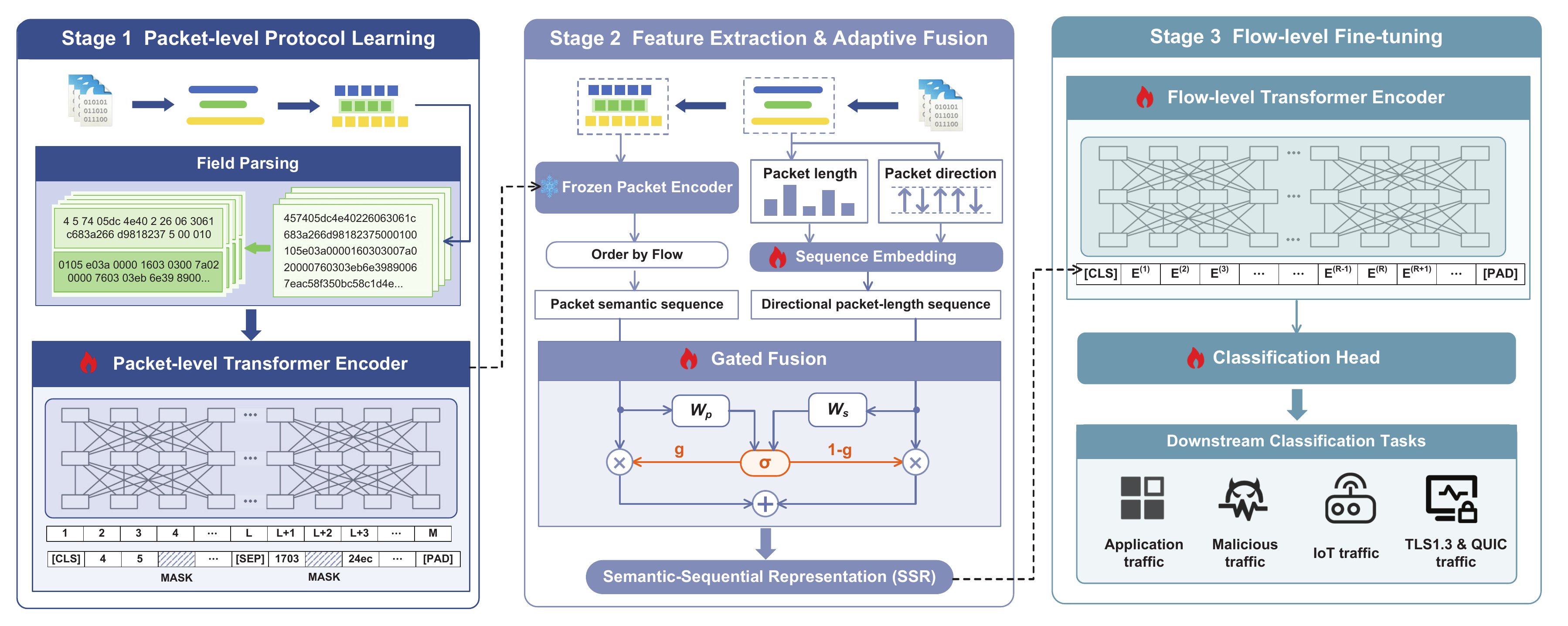}%
\caption{The overview of FLARE. The framework comprises packet-level protocol learning through field-aware masked prediction, extraction and adaptive fusion of reusable protocol-semantic representations and directional packet-length sequence, and flow-level fine-tuning for diverse traffic classification tasks.}
\label{fig1}
\end{figure*}

\section{Methodology}
\label{sec:method}

As summarized in Table~\ref{tab:field-visibility}, the observability of application-layer information, protocol-semantic metadata, and behavioral metadata varies across protection profiles. Motivated by this observation, FLARE learns parsable protocol information and observable behavioral metadata through separate representation pathways and then adaptively fuses the resulting representations for downstream classification. The overall architecture of FLARE is illustrated in Figure~\ref{fig1}. It comprises three stages: packet-level protocol learning, feature extraction and adaptive fusion, and flow-level fine-tuning.

\textbf{Packet-level Protocol Learning.} FLARE parses each packet into structured protocol fields and divides the remaining unstructured bytes into fixed-size units. A BERT-based packet encoder is then pre-trained using Masked Field Prediction (MFP) to learn field-aware packet representations from unlabeled traffic.

\textbf{Feature Extraction and Adaptive Fusion.} The pre-trained packet encoder is frozen to extract a protocol-semantic representation for each packet. In parallel, packet length and direction are combined into a directional packet-length feature. A gated fusion module adaptively integrates the semantic and sequential representations.

\textbf{Flow-level Fine-tuning.} During downstream fine-tuning, the packet-level encoder remains frozen, while the gated fusion module, flow-level encoder, and task-specific classifier are jointly optimized using labeled data. The flow-level encoder captures inter-packet dependencies and produces a task-specific flow representation for classification.

\subsection{Packet-level Protocol Learning}
As detailed in \S\ref{sec:fp} and \S\ref{sec:mfp}, each packet header is parsed into structured protocol fields and tokenized using field-aligned units. A BERT-based encoder, referred to as \textbf{PacketEncoder}, is pre-trained using the Masked Field Prediction (MFP) objective. This design encourages the encoder to capture field-aware protocol semantics and contextual dependencies among protocol fields and packet content, providing a structured semantic representation for subsequent flow-level modeling.

\subsubsection{Field Parsing}
\label{sec:fp}

In this paper, we use the term protocol semantics to refer to the meanings that protocol specifications assign to header and control fields, such as the IP and TCP/UDP header fields, the TLS record and handshake headers, and the public QUIC header fields, together with the dependencies among them. Content carried in the payload, such as application data, is not regarded as protocol semantics. As shown in Table~\ref{tab:field-visibility}, although stronger protection increasingly conceals application-layer content and service-identifying handshake metadata, these header fields remain largely observable and parsable, so protocol semantics remain available even under TLS~1.3 and QUIC. We therefore use Scapy~\cite{scapy} to dissect each packet into its constituent fields across the supported protocol layers. Each parsed field is treated as a semantic unit, allowing the tokenization process to retain field boundaries and the ordered structure of the parsed protocol elements.

During parsing, identity-related fields, including IP addresses, port numbers, and Server Name Indication (SNI), are excluded to prevent the model from exploiting endpoint-specific or service-specific shortcuts.

For each packet, we retain at most the first $B$ bytes for field parsing and tokenization, where $B$ is set to 80 by default. This byte budget typically covers the network-layer and transport-layer headers, parsable protocol information near the beginning of the packet, and a limited portion of the payload. Retaining a limited payload prefix allows the encoder to exploit application-level information when they remain observable in plaintext or partially protected traffic, while reducing the risk of excessive reliance on payload content. Bytes within the retained prefix that cannot be parsed as structured protocol fields, such as encrypted payload bytes, are divided into fixed-size 2-byte units.

After field extraction and byte segmentation, the resulting units are organized into a unified vocabulary using Byte Pair Encoding (BPE). The BPE vocabulary is learned from protocol-field values and segmented byte units by iteratively merging frequently occurring adjacent byte pairs. This process produces compact token units that reflect local byte-level regularities in both protocol-field values and the remaining packet bytes.

Following standard masked language modeling conventions, four special tokens are introduced. A [\textit{CLS}] token is prepended to each sequence, and its final hidden state is used as the aggregate packet representation. A [\textit{MASK}] token designates positions selected for the MFP task. A [\textit{PAD}] token is used to align sequences to a fixed length, and an [\textit{UNK}] token represents values that are not covered by the learned vocabulary.

\subsubsection{Masked Field Prediction}
\label{sec:mfp}
We instantiate PacketEncoder and pre-train it on unlabeled traffic using the Masked Field Prediction (MFP) task. Its input is the field-structured packet sequence produced by \S\ref{sec:fp}.

Formally, given the token sequence of a packet $\mathbf{X} = [[\textit{CLS}], t_1, t_2, \ldots, t_n]$, we randomly select \(15\%\) of token positions for prediction. For each selected position, the original token is replaced with [\textit{MASK}], replaced with a randomly sampled vocabulary token, or left unchanged with probabilities of \(80\%\), \(10\%\), and \(10\%\), respectively. PacketEncoder is trained to reconstruct the original tokens from the corrupted sequence by minimizing
\begin{equation}
    \mathcal{L}_{\mathrm{MFP}} = -\frac{1}{|\mathcal{M}|}\sum_{i \in \mathcal{M}} \log P(t_i \mid \hat{\mathbf{X}}),
\end{equation}
where $\mathcal{M}$ denotes the set of masked positions, $i$ indexes a selected position, and $\hat{\mathbf{X}}$ is the corrupted input sequence.

MFP adopts the standard masked prediction objective, but applies it to a sequence constructed from parsed protocol fields and segmented packet contents rather than to an undifferentiated byte stream. Each prediction target therefore corresponds to a protocol field or part of one, and its context follows the protocol structure of the packet. This field-aligned construction introduces an explicit structural inductive bias, encouraging the model to exploit relationships among protocol elements when reconstructing masked tokens. For example, the prediction of a masked TLS record-type token may depend on adjacent version and length fields.

For each packet, the final hidden state at the [\textit{CLS}] position serves as the semantic representation $\mathbf{h}_{\mathrm{p}} \in \mathbb{R}^{d}$. This representation encodes field-aware protocol semantics, including dependencies among structured protocol fields, together with contextualized information from segmented byte tokens.

\subsection{Feature Extraction and Adaptive Fusion}
Packet-level protocol learning produces a semantic representation for each packet by encoding structured field information and packet-content patterns. Directional packet-length sequences provide a distinct source of discriminative information by characterizing the ordered variation of packet lengths and transmission directions.

For each packet, directional packet-length is defined as the product of its packet length and a signed direction indicator. This feature remains observable under payload encryption. A gated fusion module integrates the protocol-semantic and directional packet-length representations at the packet level. The fused packet representations are subsequently arranged in arrival order and passed to the flow-level fine-tuning stage. The feature extraction and fusion procedures are described in \S\ref{sec:extraction} and \S\ref{sec:fusion}, respectively.

\subsubsection{Heterogeneous Feature Extraction}
\label{sec:extraction}

\paragraph{\textbf{Packet Semantic Sequence.}}

After pre-training, the parameters of PacketEncoder are frozen throughout downstream training. Freezing PacketEncoder preserves the protocol knowledge acquired from unlabeled traffic and reduces the number of trainable parameters during task-specific adaptation.

For the \(i\)-th packet, the tokenized packet sequence is fed into the frozen PacketEncoder, and the final hidden state at the [\textit{CLS}] position is extracted as $\mathbf{h}_{\mathrm{p}} \in \mathbb{R}^d$.

For a flow containing $N$ selected packets, the semantic representations are arranged according to packet arrival:
\begin{equation}
    \mathbf{H}_{\mathrm{sem}} = \left[\mathbf{h}_{\mathrm{p}}^{(1)}, \mathbf{h}_{\mathrm{p}}^{(2)}, \ldots, \mathbf{h}_{\mathrm{p}}^{(N)}\right] \in \mathbb{R}^{N \times d}.
\end{equation}

\paragraph{\textbf{Directional Packet-length Sequence.}}
As shown in Table~\ref{tab:field-visibility}, packet length, direction, and order remain observable across protection profiles. We therefore construct directional packet-length features to capture flow-level behavioral information.

For the $i$-th packet, the transmission direction is encoded as $d_i \in \{+1, -1\}$, where $+1$ denotes the client-to-server direction and $-1$ denotes the server-to-client direction. Given its packet length $l_i$, the directional packet-length is defined as
\begin{equation}
    \tilde{l}_i = d_i \cdot l_i.
\end{equation}

The resulting scalar is mapped to a $d$-dimensional directional packet-length representation through a learnable linear projection:
\begin{equation}
    \mathbf{h}_{\mathrm{dir}}^{(i)} = W_e \tilde{l}_i + b_e,
\end{equation}
where $\mathbf{h}_{\mathrm{dir}}^{(i)} \in \mathbb{R}^d$ denotes the directional packet-length representation of the \(i\)-th packet. This projection enables direct feature-wise fusion with \(\mathbf{h}_{\mathrm{p}}^{(i)}\).

For a flow containing $N$ selected packets, the sequential representations are arranged according to packet arrival:
\begin{equation}
    \mathbf{H}_{\mathrm{seq}} = \left[\mathbf{h}_{\mathrm{dir}}^{(1)}, \mathbf{h}_{\mathrm{dir}}^{(2)}, \ldots, \mathbf{h}_{\mathrm{dir}}^{(N)}\right] \in \mathbb{R}^{N \times d}.
\end{equation}

\subsubsection{Packet-level Gated Fusion}
\label{sec:fusion}

For the \(i\)-th packet, the protocol-semantic representation \(\mathbf{h}_{\mathrm{p}}^{(i)}\) and the directional packet-length representation \(\mathbf{h}_{\mathrm{dir}}^{(i)}\) are integrated by a gated fusion module. The concatenated representations are used to generate a feature-wise gate, which assigns a weight to each feature dimension of the two representations:

\begin{gather}
    \mathbf{g}^{(i)} = \sigma\left(\mathbf{W}_{g} \left[\mathbf{h}_{\mathrm{p}}^{(i)};
    \mathbf{h}_{\mathrm{dir}}^{(i)}\right] + \mathbf{b}_{g}\right), \label{eq:gating}\\
    \mathbf{E}^{(i)} = \mathbf{g}^{(i)} \odot \mathbf{h}_{\mathrm{p}}^{(i)} + \left(\mathbf{1}-\mathbf{g}^{(i)}\right) \odot \mathbf{h}_{\mathrm{dir}}^{(i)},
    \label{eq:gated-fusion}
\end{gather}
where $\sigma$ denotes the sigmoid function and $\odot$ denotes element-wise multiplication.

The resulting $\mathbf{E}^{(i)} \in \mathbb{R}^{d}$ integrates protocol-semantic information with directional packet-length information. Because the gate is conditioned on both input representations, their feature-wise contributions can vary across packets.

The fused packet representations are arranged in packet-arrival order to construct a flow-level sequence:
\begin{equation}
    \mathbf{H}_{\mathrm{SSR}} = \left[\mathbf{E}^{(1)}, \mathbf{E}^{(2)}, \ldots, \mathbf{E}^{(N)}\right] \in \mathbb{R}^{N \times d},
\end{equation}
We refer to $\mathbf{H}_{\mathrm{SSR}}$ as the Semantic-Sequential Representation (SSR). SSR retains both information sources at the packet level while preserving packet order for subsequent flow-level modeling.

\subsection{Flow-level Fine-tuning}

Given the Semantic-Sequential Representation \(\mathbf{H}_{\mathrm{SSR}}\), a learnable [\textit{CLS}] vector \(\mathbf{c} \in \mathbb{R}^{d}\) is prepended to the sequence:
\begin{equation}
    \mathbf{Z} = \left[ \mathbf{c}; \mathbf{H}_{\mathrm{SSR}} \right] \in \mathbb{R}^{(N+1) \times d}.
\end{equation}

The resulting sequence is fed into a BERT-based flow-level Transformer encoder, referred to as \textbf{FlowEncoder}. FlowEncoder follows the same encoder architecture as PacketEncoder but is independently initialized and optimized for each downstream task.

The final hidden state corresponding to the [\textit{CLS}] position is used as the flow-level representation $\mathbf{h}_{\text{flow}} \in \mathbb{R}^d$, which summarizes the packet-level information in SSR and the dependencies among packets in the flow.

FLARE optimizes FlowEncoder using downstream supervision and does not introduce an additional flow-level pre-training objective. The protocol-semantic representations are supplied by the frozen PacketEncoder, while task-relevant patterns in the directional packet-length features and their interactions with protocol-semantic information are learned during fine-tuning.

A task-specific classification head maps $\mathbf{h}_{\text{flow}}$ to a probability distribution over the target classes:
\begin{equation}
    \hat{\mathbf{p}} = \operatorname{Softmax}
    \left(\mathbf{W}_{2}\operatorname{GELU}
    \left(\mathbf{W}_{1}\,\mathbf{h}_{\mathrm{flow}} + \mathbf{b}_{1}\right) + \mathbf{b}_{2} \right),
\end{equation}
where $\mathbf{W}_{1}$, $\mathbf{W}_{2}$, $\mathbf{b}_{1}$, and $\mathbf{b}_{2}$ are learnable parameters of the classification head.

The model is trained by minimizing the cross-entropy loss
\begin{equation}
    \mathcal{L}_{\mathrm{cls}} = -\sum_{k=1}^{K} y_k \log \hat{p}_k,
\end{equation}
where $K$ denotes the number of target classes and $y_k$ is the ground-truth label indicator for class $k$.

During fine-tuning, the gated fusion module, FlowEncoder, and classification head are jointly optimized, whereas PacketEncoder remains frozen. Consequently, the fusion weights and flow-level representations are adapted to the supervision provided by each downstream task. This design enables FLARE to adapt effectively across diverse downstream tasks, including application classification, malicious traffic detection, website fingerprinting, and encrypted network service classification.

\section{Experiments}
\label{sec:experiments}
In this section, we describe the experiment setup and conduct a set of experiments to investigate the following Research Questions (RQ):
\begin{itemize}
    \item \textbf{RQ1:} Does FLARE achieve superior classification performance over representative baselines across diverse encrypted-traffic classification tasks?
    \item \textbf{RQ2:} Can FLARE maintain effective classification performance under strong encryption as traffic observability becomes constrained?
    \item \textbf{RQ3:} How do the learned fusion preferences between protocol-semantic and directional packet-length representations vary across traffic scenarios?
    \item \textbf{RQ4:} How does each key component contribute to the classification performance of FLARE?
    \item \textbf{RQ5:} How do the number of packets retained per flow and the number of labeled flows per class affect the classification performance of FLARE?
\end{itemize}

\subsection{Experiment Setup}
\subsubsection{Datasets}

\paragraph{\textbf{Pre-training Datasets.}}
For pre-training, we use ISCX-NonVPN~\cite{draper2016characterization}, CIRA-CIC-DoHBrw-2020~\cite{montazerishatoori2020detection}, and the attack traffic from CIC-IoT-Dataset-2022~\cite{dadkhah2022towards} as the pre-training corpus, totaling approximately 50~GB of traffic data. Since packet-level pre-training is designed to capture the semantics and structural organization of protocol fields, we deliberately construct the pre-training corpus using packets from heterogeneous protocols. The resulting corpus covers HTTP, TLS~1.2, TLS~1.3, FTP, SMTP, MQTT, CoAP, and DoH-related traffic, as well as diverse attack categories, thereby providing sufficient protocol and traffic diversity for effective unsupervised pre-training. Detailed statistics for each dataset are provided in Table~\ref{tab:pre-training datasets}.

\begin{table}[t]
\centering
\caption{Statistical details of the pre-training datasets.}
\footnotesize
\setlength{\tabcolsep}{4pt}
\begin{tabularx}{\columnwidth}{@{}l c >{\raggedright\arraybackslash}X@{}}
\toprule
Dataset & Size & Protocol \\
\midrule
ISCX-NonVPN          & 23 GB  & HTTP, TLS~1.2, FTP, SMTP \\
CIC-IoT-2022-Attacks & 5.5 GB & HTTP, MQTT, CoAP \\
DoH-Malicious        & 24 GB  & TLS~1.2, TLS~1.3 \\
\bottomrule
\end{tabularx}
\label{tab:pre-training datasets}
\end{table}

\paragraph{\textbf{Evaluation Datasets.}}
During the fine-tuning stage, we evaluate FLARE on different datasets to assess its performance on downstream tasks. The statistical details of these datasets are summarized in Table~\ref{tab:fine-tuning datasets}.

\textbf{CSTNET-TLS~1.3 Dataset~\cite{lin2022bert}} is a TLS~1.3 encrypted traffic dataset collected from the China Science and Technology Network, targeting fine-grained application identification under modern encryption.

\textbf{ISCX-VPN Dataset~\cite{draper2016characterization}} was released by the Canadian Institute for Cybersecurity in 2016, containing traffic from VPN connections. ISCX-VPN-App targets application-level identification, and ISCX-VPN-Service operates at a coarser granularity, grouping traffic into 6 service-type categories, including Chat, Email, FileTransfer, P2P, Streaming, and VoIP.

\textbf{UNSW-TON-IoT Dataset~\cite{moustafa2021new}} provides heterogeneous IoT network attack scenarios with 10 classes covering normal traffic and various attacks.

\textbf{USTC-TFC Dataset~\cite{wang2017end}} was published by the University of Science and Technology of China for traffic classification. It contains 10 benign and 10 malware traffic categories, totaling 20 classes, and is widely adopted for both malicious traffic detection and fine-grained application classification.

\textbf{CIC-IoT-2022 Dataset~\cite{dadkhah2022towards}} was released by the Canadian Institute for Cybersecurity, with traffic collected from real IoT deployments. Because its attack traffic is included in our unlabeled pre-training corpus, we exclude these samples from downstream fine-tuning. The resulting downstream task uses only benign traffic and classifies device traffic generated under different operational states.

\textbf{CipherSpectrum Dataset~\cite{wickramasinghe2025sok}} was introduced by the University of New South Wales in 2025 to address key limitations of existing datasets. It covers 41 domain classes, each represented by traffic encrypted under the three cipher suites mandated or recommended by TLS~1.3, including AES-128-GCM, AES-256-GCM, and ChaCha20-Poly1305. The dataset is designed for web traffic classification and supports cipher-agnostic network traffic classification research.

\textbf{VisQUIC Dataset~\cite{gahtan2024exploring}} is a large-scale dataset for encrypted QUIC traffic analysis. Samples are labeled by their corresponding website or server, and the dataset spans multiple QUIC implementations, including Chromium QUIC, Facebook mvfst, and Cloudflare quiche.

\begin{table*}[t]
  \centering
  \caption{
    The statistical details and collection environments of the
    fine-tuning datasets. WF denotes website fingerprinting.
  }
  \label{tab:fine-tuning datasets}

  \footnotesize
  \setlength{\tabcolsep}{3.5pt}
  \renewcommand{\arraystretch}{1.2}
  \renewcommand{\tabularxcolumn}[1]{m{#1}}

  \begin{tabularx}{\textwidth}{
    @{}
    c
    >{\centering\arraybackslash}m{2.9cm}
    >{\raggedright\arraybackslash}X
    c c c
    @{}
  }
    \toprule
    Dataset
    & Tasks
    & \multicolumn{1}{c}{Collection Environment}
    & Period
    & Flows
    & Labels \\
    \midrule

    ISCX-VPN-Service
    & VPN-based Service Type Identification
    & University lab network (UNB); researcher-generated sessions over an external VPN service
    & 2016
    & 1369
    & 6 \\

    ISCX-VPN-App
    & VPN-based Application Fingerprinting
    & University lab network (UNB); researcher-generated sessions over an external VPN service
    & 2016
    & 1358
    & 12 \\

    USTC-TFC
    & Malware Detection
    & Malware: real-network captures (CTU); benign: emulated by IXIA BPS
    & 2011--2016
    & 5202
    & 20 \\

    CIC-IoT-2022
    & IoT Detection
    & IoT lab network (CIC, about 60 devices); controlled device-activity experiments
    & 2022
    & 2000
    & 5 \\

    UNSW-TON-IoT
    & IoT Malware Detection
    & IoT/IIoT testbed (UNSW Canberra Cyber Range); emulated normal and attack traffic
    & 2019
    & 4804
    & 10 \\

    \midrule

    CSTNET-TLS~1.3
    & TLS 1.3-based WF
    & National research and education backbone (CSTNET); passively captured traffic
    & 2021
    & 7091
    & 51 \\

    CipherSpectrum
    & TLS 1.3-based WF (Multi-Cipher Suites)
    & Controlled client-side collection; automated Chromium/Firefox browsing with three TLS 1.3 AEAD cipher suites
    & 2024
    & 20500
    & 41 \\

    VisQUIC
    & QUIC-based WF
    & Two residential networks on different continents; automated Headless Chrome browsing around the clock
    & 2023--2024
    & 8000
    & 16 \\

    \bottomrule
  \end{tabularx}
\end{table*}

\subsubsection{Compared Methods}
To evaluate the performance of FLARE, we compare it with representative baselines from three methodological categories that broadly cover existing approaches to network traffic classification. \textbf{(i) Traditional Methods:} AppScanner~\cite{taylor2016appscanner} and BIND~\cite{al2016adaptive}.
\textbf{(ii) Deep Learning Methods:} LEXNet~\cite{fauvel2023lightweight}, FS-Net~\cite{liu2019fs}, DeepPacket~\cite{lotfollahi2020deep}, and Deep Fingerprinting (DF)~\cite{sirinam2018deep}. \textbf{(iii) Pre-training Methods:} ET-BERT~\cite{lin2022bert}, YaTC~\cite{zhao2023yet}, BERT-ps~\cite{yang2025robustness}, and TrafficFormer~\cite{zhou2025trafficformer}.

\subsubsection{Implementation Details}
All experiments are conducted on a server equipped with 8 NVIDIA GeForce RTX 4090 GPUs, running Python 3.9.23 and PyTorch 2.6.0 (CUDA 12.4). Both PacketEncoder and the flow-level encoder follow the standard BERT architecture, consisting of 12 Transformer encoder layers with a hidden dimension of $d = 768$ and 12 attention heads. The maximum byte sequence length per packet is set to $B = 80$, and the first $N = 64$ packets are retained per flow.

\textbf{Packet-level Pre-training.} The PacketEncoder is optimized using the Adam optimizer with a learning rate of $2 \times 10^{-5}$ over 500,000 training steps. The training loss stabilizes at approximately 300,000 steps; we therefore adopt the checkpoint at step 300,000 as the final packet-level encoder for all subsequent experiments. On all evaluated datasets, network-layer and transport-layer header fields are parsed for every packet. An independent audit based on TCP reassembly shows that 99.74\% of the TLS record headers within the retained bytes are parsed, with only 6 false positives among about 2.27 million parsed headers (Appendix A).

\textbf{Flow-level Fine-tuning.} During fine-tuning, the Adam optimizer is employed with a learning rate of $3 \times 10^{-5}$ and a batch size of 32. The maximum number of training epochs is set to 80, with early stopping applied if no improvement in validation performance is observed for 10 consecutive epochs. For downstream fine-tuning and evaluation, datasets are partitioned at the flow level into training, validation, and test sets following an 8:1:1 ratio. All packets from the same flow are retained in the same partition~\cite{zhao2025sweet}, and the split is performed before sequence construction. At most 500 flows per class are sampled from the training set. The final model is selected based on validation performance, while the test set is used only for final evaluation. To reduce the influence of random initialization and data sampling, all reported results are averaged over five independent runs under the same experimental settings.

\subsubsection{Evaluation Metrics}

We evaluate classification performance using accuracy, macro-precision, macro-recall, and macro-F1. Accuracy (Acc) measures the proportion of correctly classified samples. For each class, precision measures the proportion of samples predicted as that class that are correctly classified, while recall measures the proportion of samples belonging to that class that are correctly identified. F1 is the harmonic mean of precision and recall. Macro-precision (PR), macro-recall (RC), and macro-F1 (F1) are obtained by averaging the corresponding class-wise metrics with equal weight, thereby accounting for the performance of all classes regardless of their sample sizes. Unless otherwise specified, precision, recall, and F1 refer to their macro-averaged variants.

\begin{table*}[t]
\centering
\small
\setlength{\tabcolsep}{3pt}
\caption{Comparison results on CSTNET-TLS~1.3, ISCX-VPN-Service and ISCX-VPN-App. The best and second-best results are shown in boldface and underlined, respectively.}
\begin{tabular}{lcccccccccccc}
\toprule
\multirow{2}{*}{\textbf{Methods}} & \multicolumn{4}{c}{\textbf{CSTNET-TLS~1.3}} & \multicolumn{4}{c}{\textbf{ISCX-VPN-Service}} & \multicolumn{4}{c}{\textbf{ISCX-VPN-App}}\\
\cmidrule(lr){2-5} \cmidrule(lr){6-9} \cmidrule(lr){10-13}
 & \textbf{Acc} & \textbf{PR} & \textbf{RC} & \textbf{F1} & \textbf{Acc} & \textbf{PR} & \textbf{RC} & \textbf{F1} & \textbf{Acc} & \textbf{PR} & \textbf{RC} & \textbf{F1} \\
\midrule
AppScanner & 0.2763 & 0.7976 & 0.2475 & 0.3475 & 0.6705 & 0.8203 & 0.5396 & 0.6413 & 0.5210 & 0.6580 & 0.4488 & 0.5523 \\
BIND & 0.6755 & 0.7146 & 0.6755 & 0.6695 & 0.8476 & 0.8590 & 0.8476 & 0.8477 & 0.4292 & 0.4208 & 0.4292 & 0.3954 \\
\midrule
LEXNet & 0.7860 & 0.7743 & 0.7448 & 0.7506 & 0.8261 & 0.8313 & 0.8456 & 0.8359 & 0.8095 & 0.7584 & 0.7062 & 0.7243 \\
FS-Net & 0.8399 & 0.8194 & 0.7971 & 0.7995 & 0.8455 & 0.8689 & 0.8390 & 0.8498 & 0.7919 & 0.7311 & 0.7240 & 0.7117 \\
DeepPacket & 0.6628 & 0.4808 & 0.4024 & 0.4131 & 0.8704 & 0.7341 & 0.7056 & 0.7017 & 0.8000 & 0.7521 & 0.7097 & 0.6975 \\
DF & 0.5563 & 0.5406 & 0.5316 & 0.5193 & 0.7361 & 0.7771 & 0.7300 & 0.7417 & 0.7279 & 0.5653 & 0.5500 & 0.5449 \\
\midrule
ET-BERT & 0.5642 & 0.5085 & 0.5105 & 0.4741 & 0.8889 & 0.8994 & 0.8971 & 0.8957 & \textbf{0.9265} & \textbf{0.9174} & 0.8843 & \underline{0.8806} \\
YaTC & 0.8745 & 0.8756 & 0.8745 & 0.8701 & \underline{0.9098} & 0.9080 & 0.9098 & 0.9071 & 0.8846 & 0.8703 & \underline{0.8846} & 0.8758 \\
BERT-ps & \underline{0.9140} & \underline{0.9122} & \underline{0.9088} & \underline{0.9068} & \underline{0.9270} & \underline{0.9222} & \textbf{0.9140} & \underline{0.9132} & 0.8015 & 0.6303 & 0.6422 & 0.6319 \\
TrafficFormer & 0.7601 & 0.7258 & 0.7267 & 0.7148 & 0.8810 & 0.8863 & 0.8634 & 0.8722 & 0.8095 & 0.8069 & 0.7169 & 0.7373 \\
\midrule
FLARE & \textbf{0.9456} & \textbf{0.9406} & \textbf{0.9238} & \textbf{0.9255} & \textbf{0.9301} & \textbf{0.9293} & {0.8887} & \textbf{0.9251} & \underline{0.9189} & \underline{0.9053} & \textbf{0.9118} & \textbf{0.9009}\\
\bottomrule
\end{tabular}
\label{tab:ce1}
\end{table*}

\begin{table*}
\centering
\small
\setlength{\tabcolsep}{3pt}
\caption{Comparison results on UNSW-TON-IoT, USTC-TFC and CIC-IoT-2022. The best and second-best results are shown in boldface and underlined, respectively.}
\begin{tabular}{lcccccccccccc}
\toprule
\multirow{2}{*}{\textbf{Methods}} & \multicolumn{4}{c}{\textbf{UNSW-TON-IoT}} & \multicolumn{4}{c}{\textbf{USTC-TFC}} & \multicolumn{4}{c}{\textbf{CIC-IoT-2022}} \\
\cmidrule(lr){2-5} \cmidrule(lr){6-9} \cmidrule(lr){10-13}
 & \textbf{Acc} & \textbf{PR} & \textbf{RC} & \textbf{F1} & \textbf{Acc} & \textbf{PR} & \textbf{RC} & \textbf{F1} & \textbf{Acc} & \textbf{PR} & \textbf{RC} & \textbf{F1} \\
\midrule
AppScanner & 0.7040 & 0.7128 & 0.7040 & 0.7000 & 0.4532 & 0.8634 & 0.5946 & 0.6430 & 0.2125 & 0.4237 & 0.4610 & 0.3535 \\
BIND & 0.7944 & 0.8072 & 0.7944 & 0.7916 & 0.8451 & 0.8595 & 0.8451 & 0.8385 & 0.7939 & 0.7937 & 0.7939 & 0.7892 \\
\midrule
LEXNet & 0.7892 & 0.7916 & 0.8026 & 0.7853 & 0.7757 & 0.8709 & 0.8386 & 0.8474 & 0.7000 & 0.7001 & 0.7022 & 0.7000 \\
FS-Net & 0.8484 & 0.8422 & 0.8626 & 0.8382 & 0.7869 & 0.8663 & 0.8360 & 0.8359 & 0.7297 & 0.7331 & 0.7349 & 0.7348 \\
DeepPacket & \underline{0.9100} & \underline{0.9126} & \underline{0.9100} & \underline{0.9094} & 0.8571 & 0.6019 & 0.5769 & 0.5435 & 0.7250 & 0.7295 & 0.7250 & 0.7258 \\
DF & 0.5832 & 0.6254 & 0.5742 & 0.5687 & 0.6799 & 0.6934 & 0.6661 & 0.6491 & 0.5823 & 0.5856 & 0.5816 & 0.5627 \\
\midrule
ET-BERT & 0.8043 & 0.8259 & 0.8040 & 0.8023 & 0.9677 & 0.9746 & 0.9660 & 0.9683 & 0.7991 & 0.8208 & 0.8034 & 0.8100 \\
YaTC & 0.8882 & 0.8913 & 0.8882 & 0.8890 & 0.9733 & 0.9734 & 0.9733 & 0.9733 & \underline{0.8525} & \underline{0.8539} & \underline{0.8525} & \textbf{0.8520}\\
BERT-ps & 0.7914 & 0.8017 & 0.7920 & 0.7915 & 0.8392 & 0.8750 & 0.8785 & 0.8757 & 0.7938 & 0.8483 & 0.7816 & 0.7976 \\
TrafficFormer & 0.8609 & 0.8946 & 0.8512 & 0.8609 & \textbf{0.9952} & \underline{0.9867} & \textbf{0.9827} & \underline{0.9845} & 0.7258 & 0.7286 & 0.7278 & 0.7278 \\
\midrule
FLARE & \textbf{0.9431} & \textbf{0.9513} & \textbf{0.9381} & \textbf{0.9407} & \underline{0.9885} & \textbf{0.9908} & \underline{0.9753} & \textbf{0.9908} & \textbf{0.8621} & \textbf{0.8722} & \textbf{0.8759} & \underline{0.8349} \\
\bottomrule
\end{tabular}
\label{tab:ce2}
\end{table*}

\subsection{Comparison Experiments (RQ1)}
\label{comparison}

We compare FLARE with various baseline methods on six public traffic datasets, covering TLS~1.3 application traffic (CSTNET-TLS~1.3), VPN-tunneled traffic (ISCX-VPN-Service, ISCX-VPN-App), malicious traffic (USTC-TFC), and IoT traffic (UNSW-TON-IoT, CIC-IoT-2022). The experimental results are presented in Table~\ref{tab:ce1} and Table~\ref{tab:ce2}.

\textbf{High F1 across multiple classification tasks.} FLARE achieves the highest macro-F1 scores on five of the six datasets, with its largest gain reaching 3.13\% over the best-performing baseline on UNSW-TON-IoT. Methods based on statistical features and machine learning obtain the weakest performance; deep learning models perform better, and pre-trained models achieve the best results among the baselines. FLARE consistently outperforms methods from all three categories across VPN traffic classification, website fingerprinting, malicious traffic detection, and IoT traffic classification.

\textbf{The discriminative value of different traffic representations varies across traffic scenarios.} The experimental results show that byte-based methods hold a clear advantage over sequence-based methods on USTC-TFC: DeepPacket outperforms FS-Net by 7.02\% in accuracy, and TrafficFormer outperforms BERT-ps by 15.6\%. This is because 98.9\% of the traffic in USTC-TFC is unencrypted~\cite{wickramasinghe2025sok}, consisting largely of plaintext HTTP, custom binary protocols, or weakly encrypted communication, which allows byte-based methods to extract highly discriminative content directly from the raw traffic. In contrast, sequence-based methods are less effective because many malware flows consist of short heartbeat or command-response exchanges, with limited variation in flow length, packet count, and interaction patterns. Such simple transmission behaviors provide weak inter-class separation for sequence modeling.

CSTNET-TLS~1.3 contains encrypted traffic protected by AEAD cipher suites, as mandated by TLS~1.3. On this dataset, the sequence-based FS-Net outperforms DeepPacket by 17.71\%, and BERT-ps outperforms TrafficFormer by 15.39\%. These results indicate that byte-based models can exploit only a small amount of residual content information when facing strongly encrypted traffic, since most payload bytes no longer carry effective semantic features for classification, leading to a marked degradation in their performance. Packet lengths and directions, however, remain observable under payload encryption. In addition, CSTNET-TLS~1.3 targets website identification. Differences in page structure, resource size, and request-response order are reflected in distinctive packet-length and direction sequences, favoring sequence-based methods.

The above analysis demonstrates that the effective discriminative information of a traffic dataset resides in different features, depending on its protocol characteristics and classification tasks: for weakly encrypted malware traffic generated by homogeneous code, the discriminative information concentrates in the byte representation, whereas for strongly encrypted website traffic with heterogeneous behaviors, it derives primarily from the sequential representation. Consequently, single-representation methods tend to excel only on datasets that favor their respective representations. FLARE therefore adaptively fuses the semantic and sequential representations, which improves generalization and enables effective classification performance across six heterogeneous datasets.

\subsection{Classification under Strong Encryption (RQ2)}

\begin{table*}[t]
\centering
\small
\setlength{\tabcolsep}{3pt}
\caption{Comparison results under strong encryption across CipherSpectrum (AES-128-GCM, AES-256-GCM, ChaCha20-Poly1305, and a mixed suite), CSTNET-TLS~1.3, and VisQUIC. The SNI field is removed for all methods, and the best result in each column is shown in bold.}
\begin{tabular}{lcccccccccccc}
\toprule
\multirow{3}{*}{\textbf{Methods}}
    & \multicolumn{8}{c}{\textbf{CipherSpectrum}}
    & \multicolumn{2}{c}{\multirow{2}{*}{\textbf{CSTNET-TLS~1.3}}}
    & \multicolumn{2}{c}{\multirow{2}{*}{\textbf{VisQUIC}}} \\
\cmidrule(lr){2-9}
    & \multicolumn{2}{c}{\textbf{aes-128-gcm}}
    & \multicolumn{2}{c}{\textbf{aes-256-gcm}}
    & \multicolumn{2}{c}{\textbf{chacha20-poly1305}}
    & \multicolumn{2}{c}{\textbf{mix}}
    & \multicolumn{2}{c}{}
    & \multicolumn{2}{c}{} \\
\cmidrule(lr){2-3} \cmidrule(lr){4-5} \cmidrule(lr){6-7} \cmidrule(lr){8-9} \cmidrule(lr){10-11} \cmidrule(lr){12-13}
 & \textbf{Acc} & \textbf{F1} & \textbf{Acc} & \textbf{F1} & \textbf{Acc} & \textbf{F1} & \textbf{Acc} & \textbf{F1} & \textbf{Acc} & \textbf{F1} & \textbf{Acc} & \textbf{F1} \\
\midrule
LEXNet & 0.8650 & 0.8637 & 0.8618 & 0.8583 & 0.8645 & 0.8643 & 0.8483 & 0.8487 & 0.7860 & 0.7506 & 0.7610 & 0.7639 \\
FS-Net & 0.8746 & 0.8735 & 0.8737 & 0.8734 & 0.8745 & 0.8737 & 0.8635 & 0.8640 & 0.8399 & 0.7995 & 0.7766 & 0.7787 \\
DeepPacket & 0.1457 & 0.1402 & - & - & - & - & 0.1561 & 0.1432 & 0.6628 & 0.4131 & 0.5969 & 0.5812 \\
DF & 0.5342 & 0.5247 & 0.5548 & 0.5471 & 0.5408 & 0.5398 & 0.5360 & 0.5291 & 0.5563 & 0.5193 & 0.7962 & 0.7907 \\
\midrule
ET-BERT & 0.4464 & 0.3708 & 0.4996 & 0.4466 & 0.4589 & 0.3903 & 0.4283 & 0.3856 & 0.5642 & 0.4741 & 0.5625 & 0.3067 \\
YaTC & 0.9610 & 0.9609 & 0.9683 & 0.9682 & 0.9695 & 0.9696 & 0.9402 & 0.9402 & 0.8745 & 0.8701 & 0.6212 & 0.6481 \\
TrafficFormer & 0.9582 & 0.9580 & 0.9717 & 0.9717 & 0.9638 & 0.9637 & 0.9389 & 0.9388 & 0.7601 & 0.7148 & 0.7262 & 0.7252 \\
FLARE & \textbf{0.9861} & \textbf{0.9862} & \textbf{0.9810} & \textbf{0.9810} & \textbf{0.9963} & \textbf{0.9963} & \textbf{0.9937} & \textbf{0.9937} & \textbf{0.9456} & \textbf{0.9255} & \textbf{0.8354} & \textbf{0.8353} \\
\bottomrule
\end{tabular}
\label{tab:se}
\end{table*}

\begin{figure*}
    \centering
    \begin{subfigure}[b]{0.24\textwidth}
        \centering
        \includegraphics[width=\textwidth]{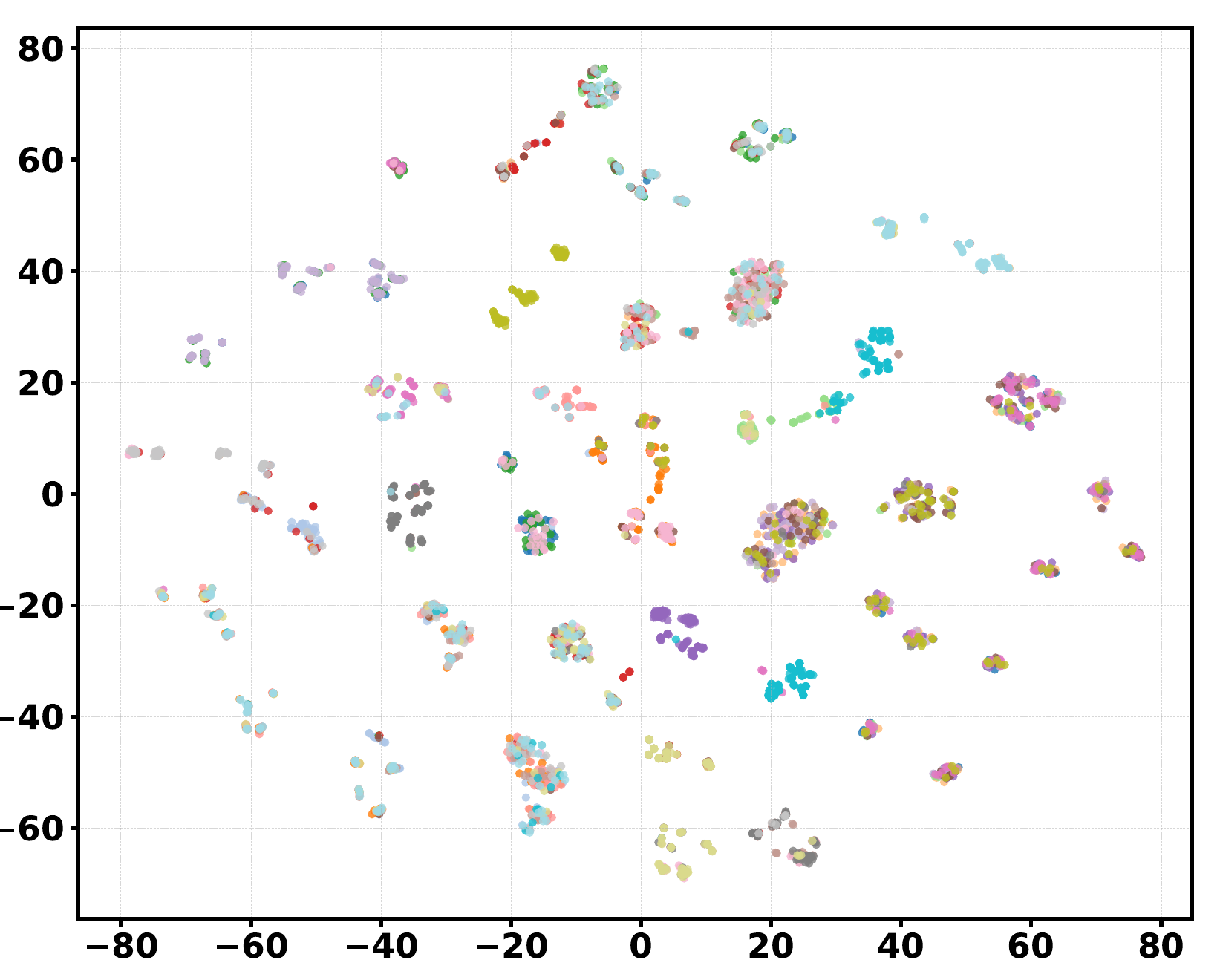}
        \caption{ET-BERT}
        \label{fig:sub1}
    \end{subfigure}
    \hfill
    \begin{subfigure}[b]{0.24\textwidth}
        \centering
        \includegraphics[width=\textwidth]{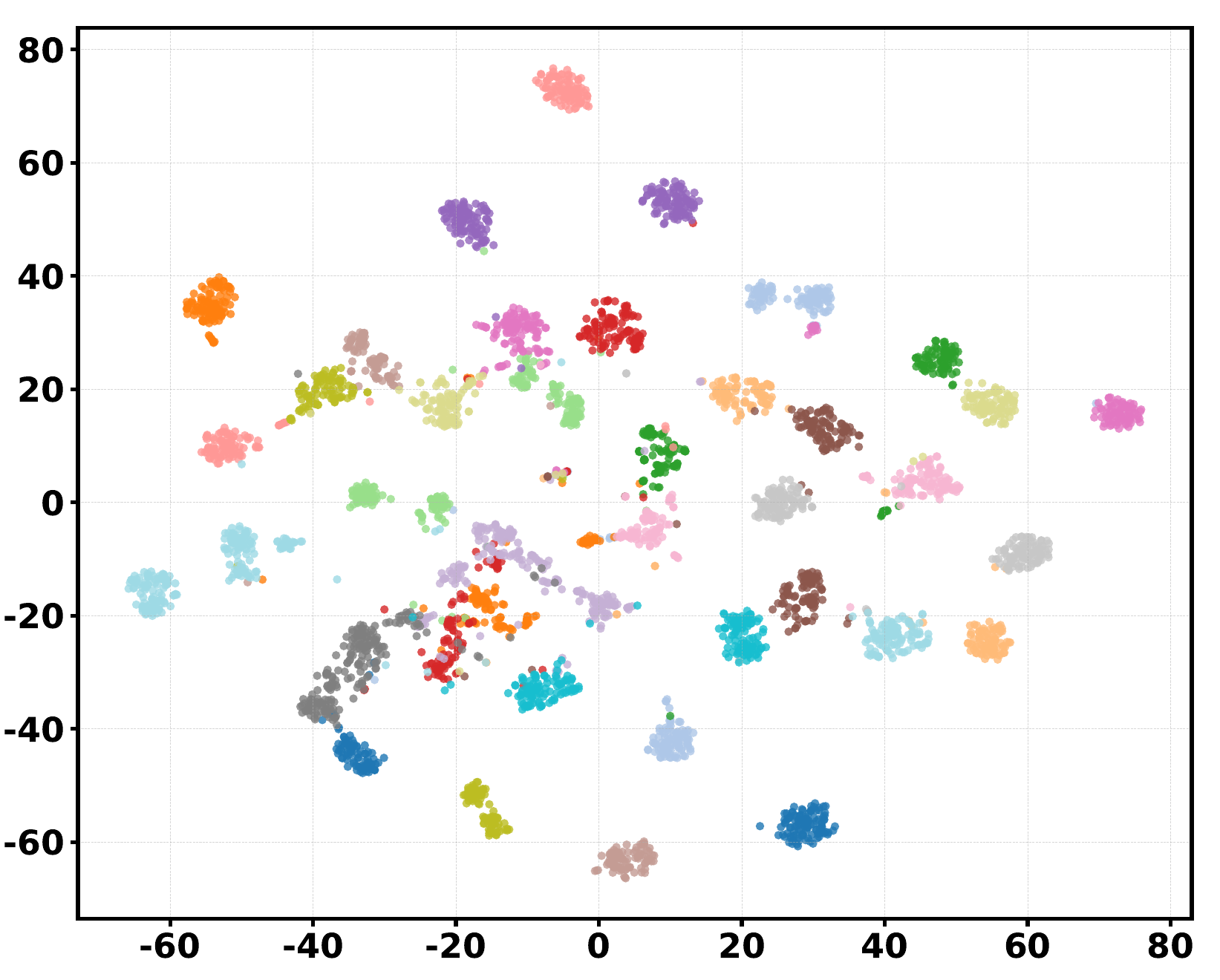}
        \caption{FS-Net}
        \label{fig:sub2}
    \end{subfigure}
    \hfill
    \begin{subfigure}[b]{0.24\textwidth}
        \centering
        \includegraphics[width=\textwidth]{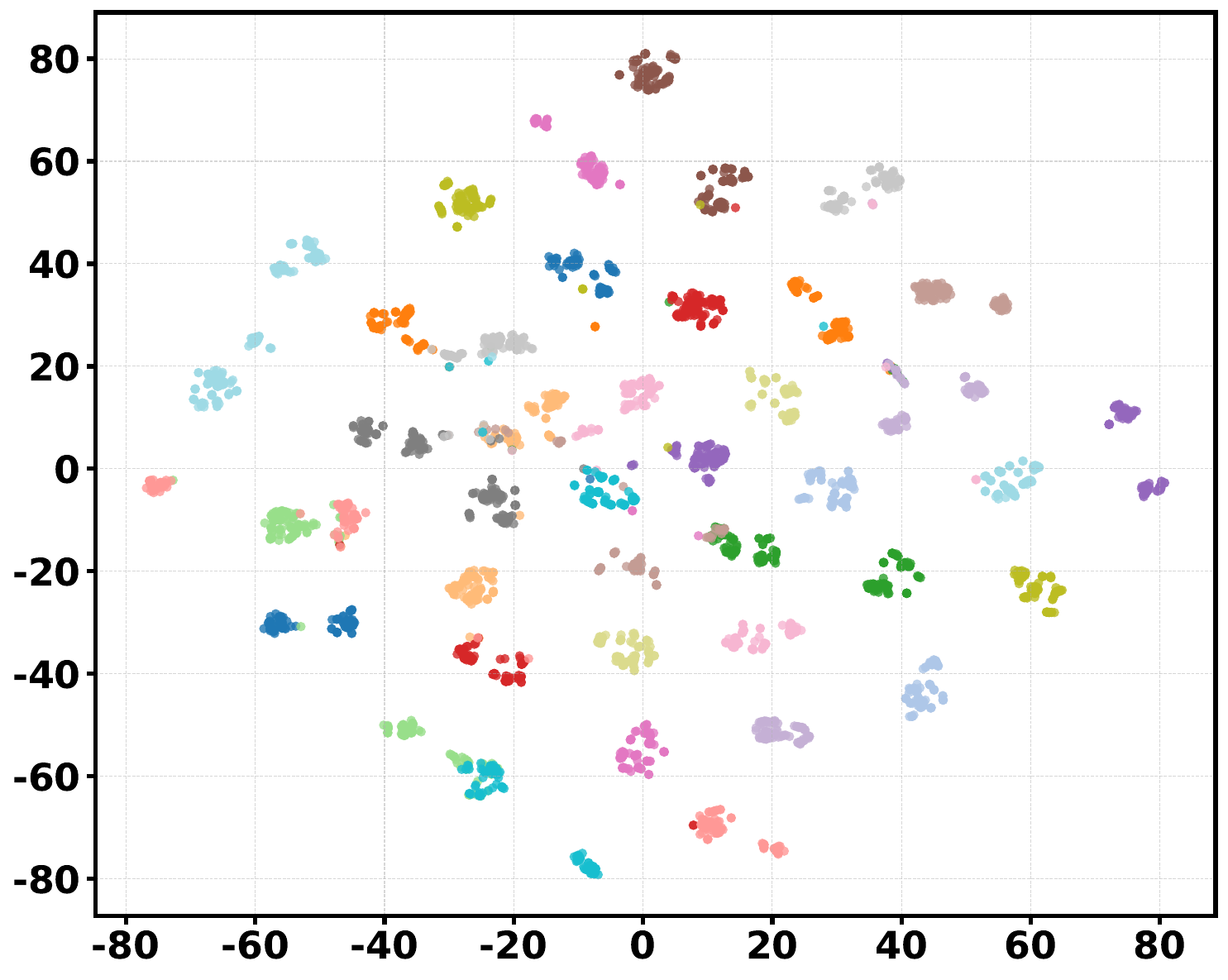}
        \caption{YaTC}
        \label{fig:sub3}
    \end{subfigure}
    \hfill
    \begin{subfigure}[b]{0.24\textwidth}
        \centering
        \includegraphics[width=\textwidth]{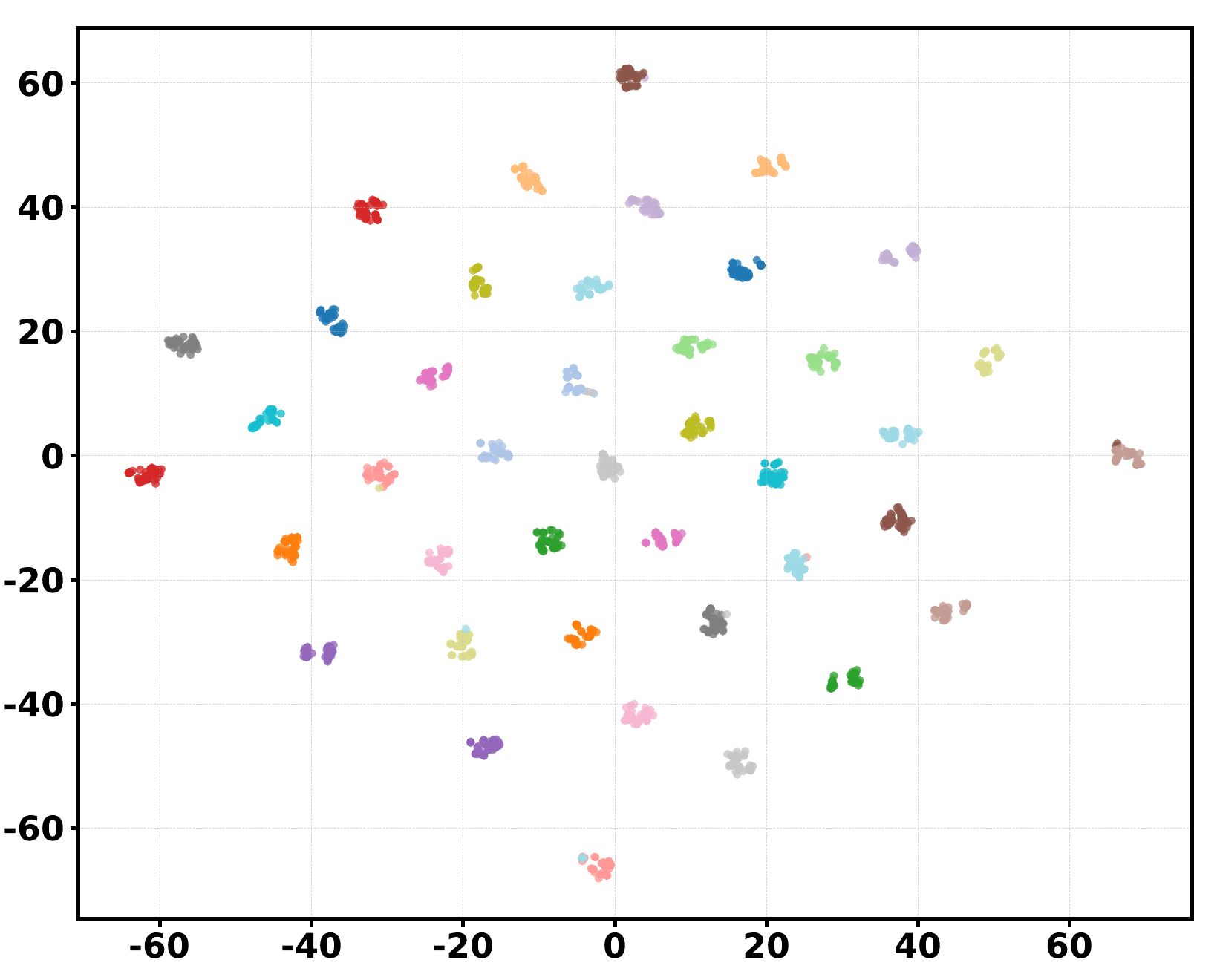}
        \caption{FLARE}
        \label{fig:sub4}
    \end{subfigure}
    \caption{The \textit{t}-SNE visualization of feature representations on CipherSpectrum-mix dataset.}
    \label{fig:overall}
\end{figure*}

\begin{table}
\centering
\small
\caption{Accuracy of FLARE under inference-time masking of shortcut-prone header fields. $\Delta$ denotes the change in accuracy, measured in percentage points, relative to the full-input setting.}
\begin{tabular}{lcccc}
\toprule
\multirow{2}{*}{\textbf{Masked fields}}
  & \multicolumn{2}{c}{\textbf{CipherSpectrum-mix}}
  & \multicolumn{2}{c}{\textbf{CSTNET-TLS~1.3}} \\
\cmidrule(lr){2-3} \cmidrule(lr){4-5}
  & \textbf{Acc} & \textbf{$\Delta$(pp)} & \textbf{Acc} & \textbf{$\Delta$(pp)} \\
\midrule
None           & 0.9937 & --      & 0.9456 & --      \\
\midrule
IP ID          & 0.9801 & \textbf{$-$1.36}  & 0.9337 & $-$1.19 \\
Checksum       & 0.9840 & $-$0.97  & 0.9589 & $+$1.33 \\
\midrule
ACK             & 0.9864 & $-$0.73  & 0.9369 & $-$0.87 \\
SEQ           & 0.9854 & $-$0.83  & 0.9298 & \textbf{$-$1.58} \\
TCP Window    & 0.9835 & $-$1.02  & 0.9305 & $-$1.51 \\
\bottomrule
\end{tabular}
\label{tab:shortcut}
\end{table}

To evaluate FLARE under strong encryption, we select three website identification datasets representing progressively more constrained protocol observability. CipherSpectrum encrypts traffic with the AEAD cipher suites mandated by TLS~1.3, covering AES-128-GCM-SHA256, AES-256-GCM-SHA384, CHACHA20-POLY1305-SHA256, and a mixed subset combining all three, which enables a fine-grained comparison across cipher suites. Since its ClientHello messages expose SNI in plaintext, we remove the SNI field to prevent identity-based shortcut learning. CSTNET-TLS~1.3 also contains TLS~1.3 traffic. For this dataset, we discard the packet containing the ClientHello message and all preceding packets in each flow, thereby removing both the plaintext SNI and additional early-connection context. VisQUIC consists of QUIC traffic, in which most transport-control information is carried in protected frames and the packet number and selected header bits are subject to header protection. Consequently, VisQUIC exposes fewer directly visible transport-layer fields than TLS-over-TCP datasets. Under these preprocessing settings, none of the evaluated methods has access to SNI information. Detailed dataset statistics are provided in Table~\ref{tab:fine-tuning datasets}.

\textbf{High accuracy under increasingly constrained observability.}
FLARE achieves the highest accuracy on all strongly encrypted datasets. Compared with the strongest pre-trained baseline on each dataset, FLARE improves the average accuracy by 2.95\% on CipherSpectrum, 7.11\% on CSTNET-TLS~1.3, and 10.92\% on VisQUIC, respectively; compared with the strongest deep learning baseline, the improvements are 11.77\%, 10.57\%, and 3.92\%, respectively. Furthermore, across the four cipher-suite subsets of CipherSpectrum, the accuracy of FLARE varies by at most 1.53\%, indicating that its classification capability is agnostic to the specific cipher suite. Notably, FLARE is the only model that consistently surpasses both all pre-trained and all deep learning baselines across the three datasets.

\textbf{Payload-based methods perform poorly under strong encryption.} ET-BERT, which relies entirely on payload data, attains average accuracies of only 45.83\% on CipherSpectrum (averaged across its four cipher-suite subsets), 56.42\% on CSTNET-TLS~1.3, and 56.25\% on VisQUIC, a marked decline from its performance on weakly encrypted public datasets in \S\ref{comparison}. Under strong payload protection, byte-oriented models may rely on ciphertext-specific statistical regularities because directly interpretable application semantics are largely unavailable. In contrast, YaTC implicitly captures inter-packet information through its byte-matrix arrangement, and TrafficFormer learns the ordering relations among bursts through its pre-training tasks; both consequently retain considerably better performance on the three datasets. This disparity indicates that the degradation of pre-trained models is proportional to their reliance on payload semantics, which further corroborates the value of sequence-level information under strong encryption.

\textbf{Sequence features remain discriminative under strong encryption.} Sequence-based methods such as LEXNet and FS-Net maintain competitive and stable classification performance across the three datasets, and in certain settings surpass some pre-trained methods. Notably, on VisQUIC, the most strongly encrypted dataset, DF outperforms all pre-trained baselines, providing direct evidence that sequential features alone retain strong discriminative information. Nevertheless, sequence-based methods face an inherent performance ceiling, as they are unable to exploit the residual plaintext fields in protocol interactions. FLARE still outperforms the strongest sequence-based method by 3.92\% on VisQUIC, and this margin is attributable to its semantic representation learning by parsing the remaining header fields, together with the gating mechanism effectively arbitrating between semantic and sequential representations. These results validate the design of FLARE: field-level semantic parsing preserves the residual value of the semantic representation under strong encryption, while allowing the model to rely more on the sequential representation if semantic information diminishes, jointly enabling the best classification performance under diverse traffic classification tasks.

\textbf{Performance does not stem from shortcut fields.} To determine whether FLARE relies on shortcut-prone header fields, we individually mask the IP ID, checksum, TCP sequence number, acknowledgment number, and window size at inference time. As shown in Table~\ref{tab:shortcut}, no single-field masking operation causes a substantial performance collapse. The largest accuracy decreases are 1.36\% after masking the IP ID on CipherSpectrum-mix and 1.58\% after masking the sequence number on CSTNET-TLS~1.3. Moreover, removing the checksum improves the accuracy on CSTNET-TLS~1.3 by 1.33\%, suggesting that it is not a stable discriminative field. These results indicate that FLARE does not rely heavily on any tested field.

\textbf{FLARE learns more separable representations under strong encryption.} Figure~\ref{fig:overall} visualizes the feature representations learned by different methods on CipherSpectrum-mix. ET-BERT and YaTC exhibit noticeable inter-class overlap and scattered within-class distributions for several traffic categories. FS-Net forms several relatively compact clusters, indicating that packet-length and direction sequences retain useful behavioral information under strong encryption, although some behaviorally similar classes remain entangled. In contrast, FLARE produces clearer class boundaries with less local overlap, demonstrating improved intra-class consistency and inter-class separability. This improvement can be attributed to the joint modeling of protocol semantics from packet-header fields and observable packet-length and direction patterns. The visualization is consistent with the quantitative classification results.

\subsection{Fusion Preference Analysis (RQ3)}
\label{adaptive}
To characterize the fusion preferences learned for different downstream datasets, we analyze the feature-wise gates produced by the gated fusion module. As defined in Eqs.~\eqref{eq:gating} and~\eqref{eq:gated-fusion}, $\mathbf{g}^{(i)}(x)$ assigns a weight to each feature dimension of the protocol-semantic representation at the $i$-th packet position of flow $x$, while the directional packet-length representation receives the complementary weight $\mathbf{1}-\mathbf{g}^{(i)}(x)$.

We define the dataset-level semantic fusion preference score as the mean semantic gate allocation over all non-padding packet positions:
\begin{equation}
    \bar{g}_{\mathrm{sem}}(D) = \frac{1}{|D|} \sum_{x \in D} \frac{1}{L_{x}} \sum_{i=1}^{L_{x}} \frac{1}{d} \sum_{k=1}^{d} g_{k}^{(i)}(x),
    \label{eq:semantic-reliance}
\end{equation}
where $D$ denotes the set of flows in the dataset, $L_x$ is the number of non-padding packet positions in flow $x$, and $g_{k}^{(i)}(x)$ denotes the $k$-th component of the gate vector at the $i$-th packet position. The corresponding sequential fusion preference score is defined as
\begin{equation}
    \bar{g}_{\mathrm{seq}}(D)
    =
    1-\bar{g}_{\mathrm{sem}}(D).
\end{equation}

By construction, $\bar{g}_{\mathrm{sem}}(D), \bar{g}_{\mathrm{seq}}(D) \in [0,1]$ and $\bar{g}_{\mathrm{sem}}(D) + \bar{g}_{\mathrm{seq}}(D) = 1$. These scores summarize the average gate allocations assigned to the protocol-semantic and directional packet-length representations and serve as proxies for the fusion preference learned on dataset $D$. A higher $\bar{g}_{\mathrm{sem}}(D)$ indicates that the gated fusion module assigns greater average weight to protocol-semantic representation, whereas a higher $\bar{g}_{\mathrm{seq}}(D)$ indicates greater average weight on directional packet-length representation. Comparing these scores across datasets characterizes the task-specific fusion preferences learned during downstream fine-tuning.

\begin{figure}
\centering
\includegraphics[width=1\columnwidth]{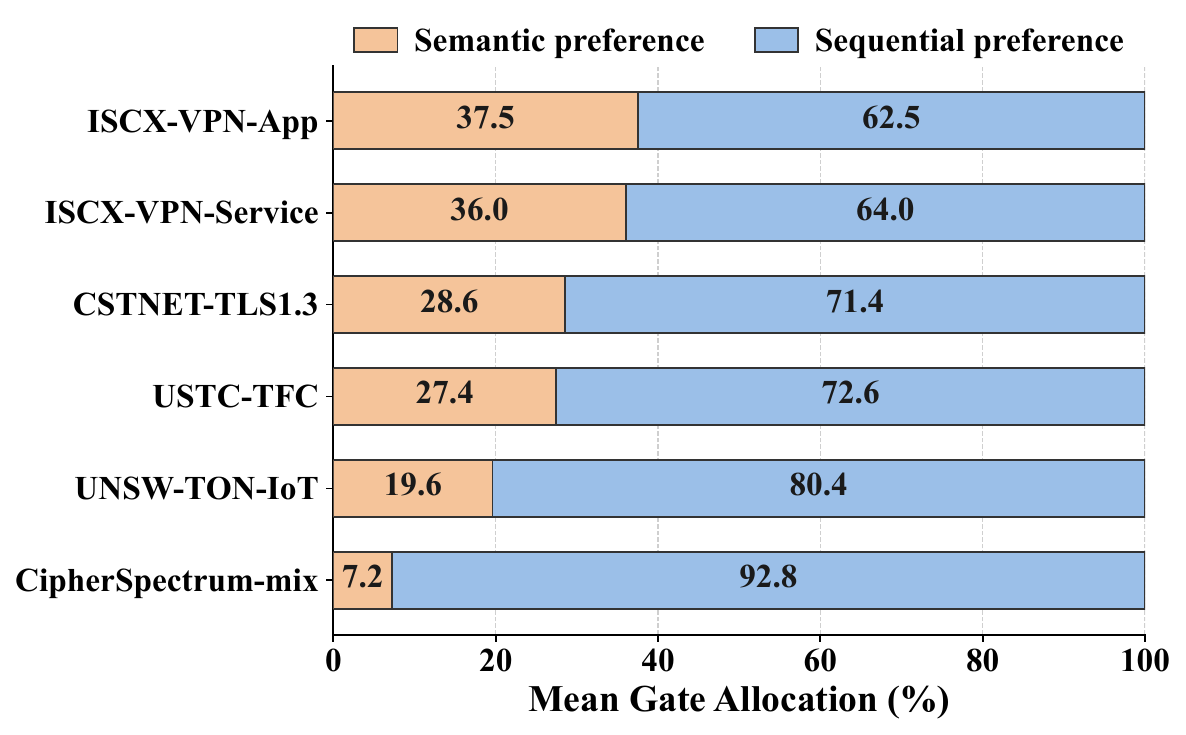}
\caption{Fusion preferences learned by FLARE for different downstream datasets. Semantic preference is quantified by the mean semantic gate $\bar{g}_{\mathrm{sem}}(D)$, while sequential preference is defined as $1-\bar{g}_{\mathrm{sem}}(D)$. These quantities serve as dataset-level proxies for the relative weighting of the two representations. Datasets are ordered by decreasing semantic preference.}
\label{fig:modality-reliance}
\end{figure}

We conduct the fusion preference analysis on CSTNET-TLS~1.3, ISCX-VPN-Service, ISCX-VPN-App, UNSW-TON-IoT, USTC-TFC, and the mixed cipher-suite subset of CipherSpectrum (CipherSpectrum-mix). The results are shown in Figure~\ref{fig:modality-reliance}.

Across the six dataset-specific fine-tuned models, the mean semantic preference ranges from 7.2\% to 37.5\%, with the corresponding sequential preference ranging from 62.5\% to 92.8\%. These differences indicate that task-specific fine-tuning yields distinct fusion preferences rather than a uniform weighting scheme across datasets. Among the evaluated datasets, ISCX-VPN-Service and ISCX-VPN-App exhibit the highest semantic preference, suggesting that the parsable protocol information retained after VPN encapsulation remains useful for these classification tasks. In contrast, CipherSpectrum-mix exhibits the lowest semantic preference, as its traffic is strongly encrypted under TLS~1.3, leaving few informative fields available, and the model consequently assigns greater weight to the sequential representation. This pattern is consistent with the effectiveness of semantic and sequential representations discussed in \S\ref{comparison}.

USTC-TFC deserves particular discussion. Although \S\ref{comparison} shows that this dataset favors byte-based classification due to its largely unencrypted payloads, FLARE deliberately restricts its semantic input to a short byte prefix and emphasizes parsed protocol fields rather than full-payload content, in order to remain effective under strong encryption and to avoid shortcut features. As a result, the semantic representation cannot exploit the signature-like payload patterns in USTC-TFC, and the model shifts most of its preference to the sequential representation. This design choice may partly explain why FLARE performs slightly below TrafficFormer on this dataset, representing a deliberate trade-off between exploiting plaintext payloads and maintaining robustness on encrypted traffic. Notably, the semantic preference on USTC-TFC still reaches 27.4\%, considerably higher than that on CipherSpectrum-mix, indicating that header fields alone provide non-negligible discriminative information even without access to payload content.

These results indicate that the gated fusion module learns distinct task-specific preferences for the protocol-semantic and directional packet-length representations. By adaptively adjusting the gate under downstream supervision, FLARE can better exploit the discriminative information available in different traffic scenarios.

\begin{figure}
\centering
\includegraphics[width=1\columnwidth]{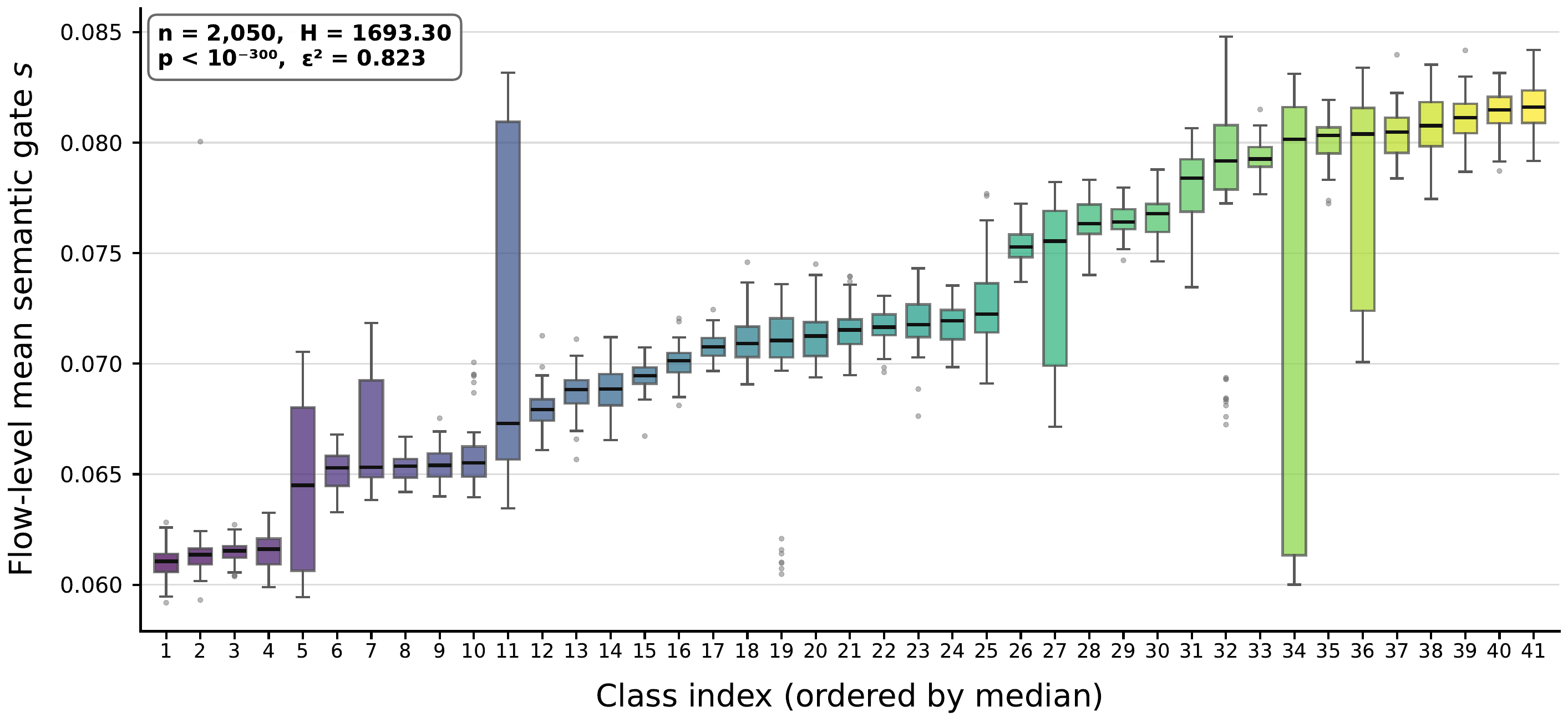}
\caption{Distribution of the mean gate value of each flow, averaged over its packets and gate dimensions, across the 41 classes of CipherSpectrum-mix. Classes are ordered by median.}
\label{fig:cm-per-class}
\end{figure}

\begin{figure}
\centering
\includegraphics[width=1\columnwidth]{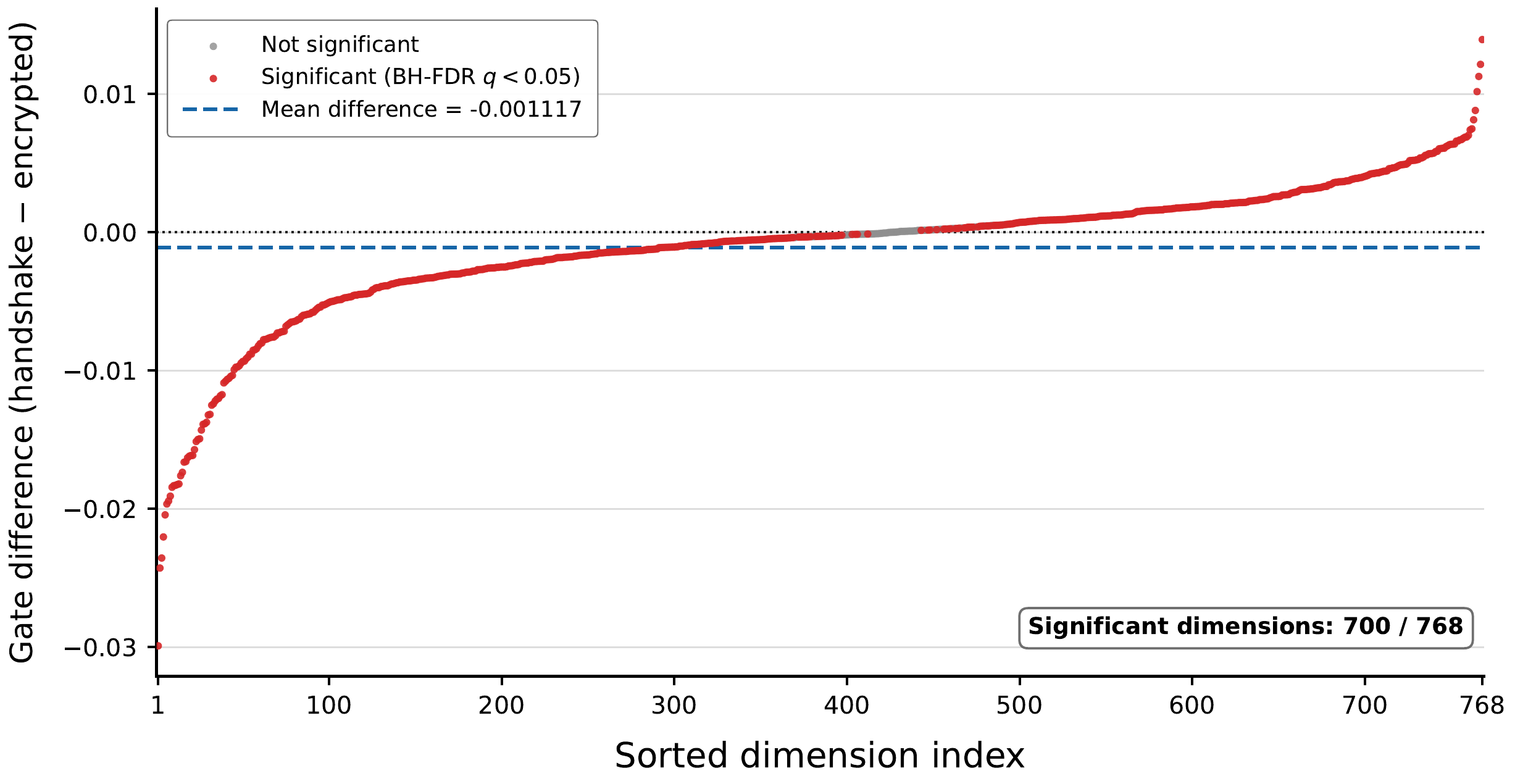}
\caption{Per-dimension difference in mean gate value between plaintext handshake records and encrypted records (handshake minus encrypted) on CipherSpectrum-mix, sorted in ascending order. Red points denote dimensions that differ significantly after Benjamini–Hochberg correction (adjusted $p<0.05$). The dashed line marks the mean difference across all dimensions, which is close to zero because positive and negative differences cancel out. Encrypted records include both the encrypted remainder of the handshake and application data.}
\label{fig:cm-per-type}
\end{figure}

\textbf{Fine-grained gate analysis.} Beyond dataset-level preferences, the gate also varies across classes and between packet types within a flow, although these variations are largely hidden in the mean gate value. On CipherSpectrum-mix, class membership explains 82.3\% of the variation in the mean gate value of a flow (Figure~\ref{fig:cm-per-class}), and a weaker class effect is observed on VisQUIC. By contrast, packet position explains less than 0.5\% of this variation, indicating that the overall balance between the two representations remains stable along a flow. The gate nevertheless changes systematically between packet types. On CipherSpectrum-mix, plaintext handshake records, which expose parsable protocol fields, and the encrypted records that follow receive almost identical mean gate values, yet 700 of the 768 gate dimensions differ significantly between them after correction for multiple comparisons (Figure~\ref{fig:cm-per-type}). Because these dimensions shift in opposite directions, their differences cancel out in the mean. Global-weight fusion uses one weight vector for all packets, and flow-level gated fusion uses one gate for all packets within a flow, so neither can express such changes between packets of the same flow. This may explain the advantage of packet-level gating in the ablation study.

\begin{table*}[t]
\centering
\caption{Ablation Study of FLARE on four datasets.}
\begin{tabular}{lcccccccc}
\toprule
\multirow{2}{*}{\textbf{Modules}} & \multicolumn{2}{c}{\textbf{CSTNET-TLS~1.3}} & \multicolumn{2}{c}{\textbf{ISCX-VPN-App}} & \multicolumn{2}{c}{\textbf{UNSW-TON-IoT}} & \multicolumn{2}{c}{\textbf{CipherSpectrum-mix}}\\
\cmidrule(lr){2-3} \cmidrule(lr){4-5} \cmidrule(lr){6-7} \cmidrule(lr){8-9}
\multicolumn{1}{c}{} & \textbf{Acc} & \textbf{F1} & \textbf{Acc} & \textbf{F1} & \textbf{Acc} & \textbf{F1} & \textbf{Acc} & \textbf{F1} \\
\midrule
w/o Semantic representation & 0.9057 & 0.8875 & 0.8311 & 0.6831 & 0.9183 & 0.9176 & 0.9777 & 0.9750 \\
Byte-level pre-training & 0.8167 & 0.7888 & 0.7905 & 0.7110 & 0.8047 & 0.8007 & 0.9707 & 0.9709 \\
\midrule
w/o Sequential representation & 0.8184 & 0.7779 & 0.7905 & 0.7085 & 0.8346 & 0.8156 & 0.8493 & 0.8475 \\
w/o Sequential feature \& length fields & 0.4529 & 0.3876 & 0.7310 & 0.7287 & 0.6629 & 0.6395 & 0.7165 & 0.7151 \\
w/o Packet length & 0.8619 & 0.8348 & 0.8311 & 0.8196 & 0.8743 & 0.8610 & 0.8946 & 0.8942 \\
w/o Packet direction & 0.9442 & \textbf{0.9341} & 0.8851 & 0.8737 & 0.9313 & 0.9290 & 0.9907 & 0.9907 \\
\midrule
Equal-weight fusion & 0.9124 & 0.8864 & 0.8176 & 0.7043 & 0.9001 & 0.8891 & 0.9917 & 0.9917 \\
Global-weight fusion & 0.9216 & 0.9038 & 0.9094 & 0.8957 & 0.9033 & 0.8977 & 0.9851 & 0.9852\\
Flow-level gated fusion & 0.9282 & 0.9051 & 0.8850 & 0.8752 & 0.9258 & 0.9211 & 0.9902 & 0.9902 \\
\midrule
\textbf{FLARE} & \textbf{0.9456} & 0.9255 & \textbf{0.9189} & \textbf{0.9009} & \textbf{0.9431} & \textbf{0.9407} & \textbf{0.9937} & \textbf{0.9937} \\
\bottomrule
\end{tabular}
\label{tab:ab}
\end{table*}

\subsection{Ablation Study (RQ4)}
\label{sec:ablation}

To validate the design of FLARE, we conduct ablation studies on four representative datasets, including CSTNET-TLS~1.3, ISCX-VPN-App, UNSW-TON-IoT, and CipherSpectrum-mix, to assess the contribution of each component. The results are presented in Table~\ref{tab:ab}.

 \textbf{Protocol semantics are effective, and field-aware pre-training preserves them.} Removing the semantic representation degrades the classification accuracy of FLARE by 1.60 percentage points on CipherSpectrum-mix, 3.99 on CSTNET-TLS~1.3, 2.48 on UNSW-TON-IoT, and 8.78 on ISCX-VPN-App. The reduction is smallest on CipherSpectrum-mix, where stronger protection leaves less informative protocol-field information available to the model. Across datasets, the size of the drop follows the fusion preferences measured in \S\ref{adaptive}: ISCX-VPN-App, which exhibits the highest semantic preference, suffers the largest drop, whereas CipherSpectrum-mix, with the lowest preference, is affected the least. Notably, even on CipherSpectrum-mix, removing the semantic representation still causes a non-negligible loss, indicating that the semantic branch provides discriminative information across all scenarios. To further isolate the effect of field-level parsing, we replace the parsed fields with raw bytes for pre-training while keeping the architecture, pre-training objective, and training procedure identical. This byte-level variant underperforms FLARE by 10.47 percentage points on average. Since the two variants differ only in how the input is constructed, this gap indicates that field-aligned input enables the encoder to learn more effective protocol-semantic representations of packets.

Since length fields encode information that overlaps with the sequential representation, we further mask the length fields of the network and transport layers in the variant without the sequential representation. Accuracy decreases on all four datasets, most notably by 36.55 percentage points on CSTNET-TLS~1.3, but remains at 71.65\%, 45.29\%, 73.10\%, and 66.29\% on CipherSpectrum-mix, CSTNET-TLS~1.3, ISCX-VPN-App, and UNSW-TON-IoT, respectively. Beyond length information, the remaining protocol fields thus provide discriminative features, even on strongly encrypted traffic.

\textbf{Sequential features are essential for classification.} Removing the sequential representation causes the largest degradation among all ablation settings, with an average accuracy drop of 12.71\% and a maximum drop of 14.44\% on CipherSpectrum-mix. Notably, CipherSpectrum-mix, which suffers the smallest loss in the semantic ablation (1.60\%), exhibits the largest loss when the sequential branch is removed (14.44\%), a contrast consistent with its dominant sequential preference measured in \S\ref{adaptive}. We further decompose the sequential features into the packet length sequence and the packet direction sequence. The length sequence encodes the volume characteristics of the transmitted content and distinguishes the functional roles of packets, such as full-sized data packets versus short control messages. The direction sequence characterizes the interaction pattern between endpoints, reflecting the request-response rhythm and the traffic asymmetry between, for instance, bulk-download services and interactive applications. Removing the length sequence reduces the average accuracy by 8.49\%, compared with 1.25\% for the direction sequence, indicating that both components contribute to classification while the length sequence plays the more prominent role. By fusing directional packet-length sequence with the semantic representation through the gating network, FLARE exploits the transmission patterns that remain discriminative regardless of encryption, which underpins its performance across diverse datasets.

\textbf{Fine-grained adaptive gated fusion is effective.}
To evaluate the fusion design, we compare FLARE with three variants of increasing adaptivity. The equal-weight variant bypasses the gating network and fuses the semantic and sequential representations with a constant weight of 0.5. The global-weight variant replaces the gate with feature-wise weights that are learned during fine-tuning but shared by all inputs, thereby allowing a dataset-specific fusion ratio. The flow-level variant computes a single gate for each flow from the pooled packet representations. As shown in Table~\ref{tab:ab}, FLARE achieves the highest accuracy on all four datasets. The equal-weight variant shows the largest degradation, reaching 10.13 percentage points on ISCX-VPN-App. The global-weight and flow-level variants remain 0.86 to 3.98 and 0.35 to 3.39 percentage points below FLARE, respectively. These results suggest that computing the fusion weights separately for each packet, conditioned on its own representations, is more effective than fusion that adapts only at the dataset or flow level.

\subsection{Sensitivity Analysis (RQ5)}

\begin{figure}
    \centering

    \begin{subfigure}[t]{0.48\columnwidth}
        \centering
        \includegraphics[width=\linewidth]
            {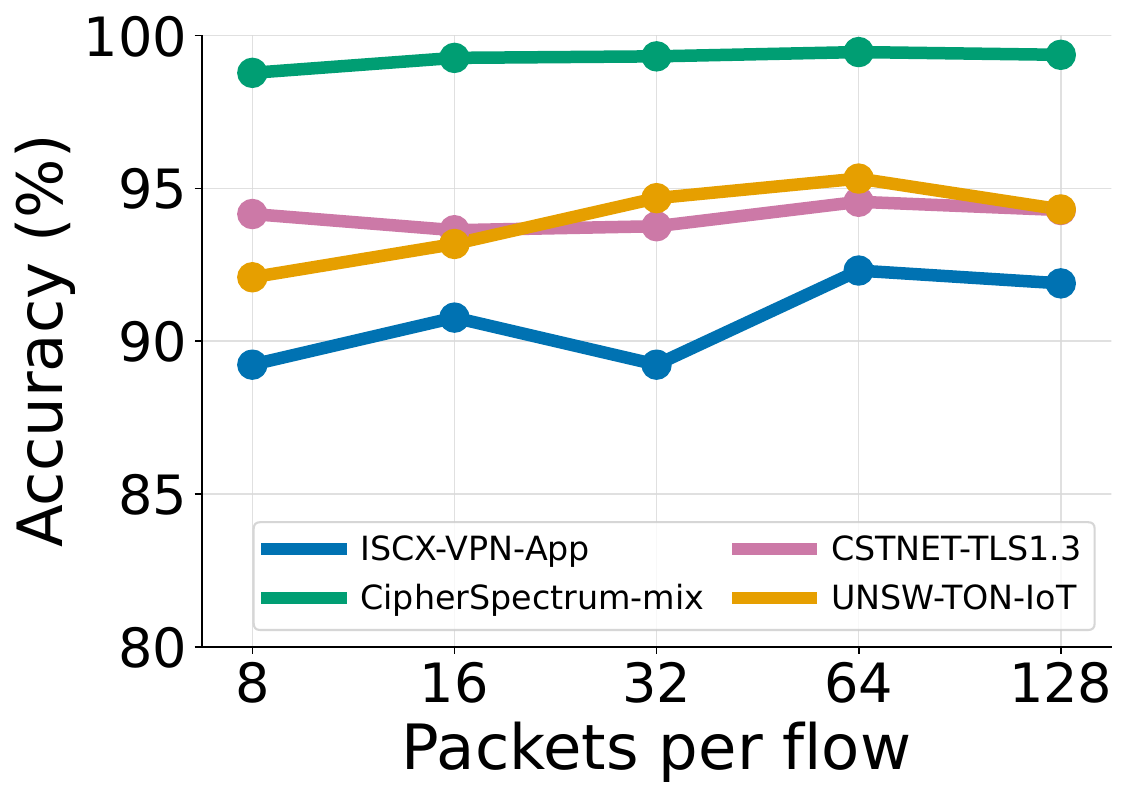}
        \caption{Number of packets per flow}
        \label{fig:sens-packets}
    \end{subfigure}
    \hfill
    \begin{subfigure}[t]{0.48\columnwidth}
        \centering
        \includegraphics[width=\linewidth]
            {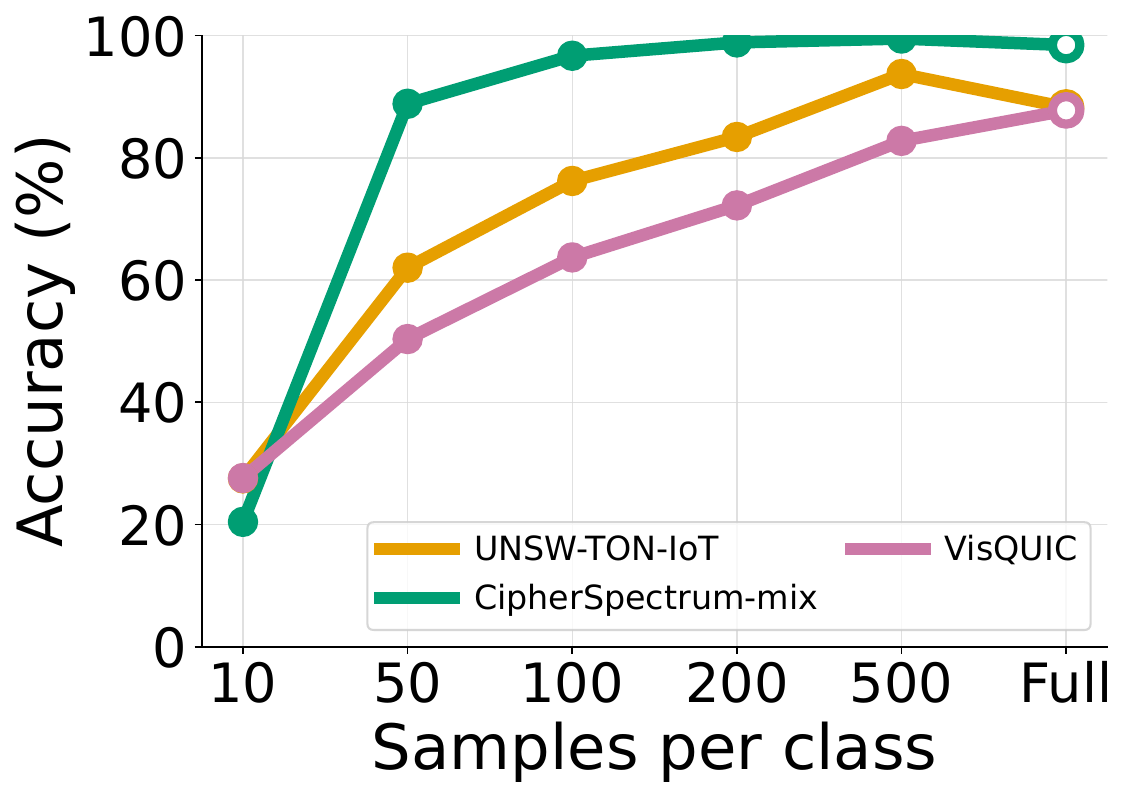}
        \caption{Number of flows per class}
        \label{fig:sens-samples}
    \end{subfigure}

    \caption{Sensitivity of FLARE to (a) the number of packets per flow and (b) the number of flows per class. ``Full'' denotes all remaining flows after constructing the test set.}
    \label{fig:param-sensitivity}
    \vspace{-3mm}
\end{figure}

We examine the sensitivity of FLARE to two parameters that determine the amount of input and supervision: the number of packets retained per flow and the number of labeled training flows per class. The results are shown in Figure~\ref{fig:sens-packets} and Figure~\ref{fig:sens-samples}, respectively.

\textbf{How many packets per flow are needed?} We vary the number of packets retained per flow as $N \in \{8,16,32,64,128\}$ on CSTNET-TLS~1.3, ISCX-VPN-App, UNSW-TON-IoT, and CipherSpectrum-mix, while fixing the per-packet input length to $B$, consistent with the pre-training configuration.

Despite minor fluctuations at smaller packet counts, accuracy generally increases with $N$ and reaches its highest or near-highest level at $N=64$ on all four datasets, whereas increasing $N$ to 128 yields a slight decrease, suggesting that later packets contribute limited additional information and may introduce redundant patterns. The datasets also differ in their sensitivity. FLARE already achieves high accuracy with the first eight packets on CipherSpectrum-mix and CSTNET-TLS~1.3, indicating that much of the useful information is concentrated in the early stage of a flow, whereas ISCX-VPN-App and UNSW-TON-IoT benefit more from longer sequences. Considering validation performance and input overhead, we set $N=64$ as the default.

\textbf{How many training flows per class are needed?} To examine the effect of the 500-flow bound on the training set, we conduct this analysis on UNSW-TON-IoT, CipherSpectrum-mix, and VisQUIC, which cover IoT attack traffic, TLS~1.3 traffic, and QUIC traffic, respectively. In these datasets, most classes contain more than 500 flows, so the bound determines how much of the available labeled data is used; in the remaining datasets, most classes contain fewer than 500 flows, and the bound has little effect. For each dataset, we fix a class-balanced test set of 100 flows per class and vary the number of training flows per class as $M \in \{10, 50, 100, 200, 500, \text{Full}\}$, where Full denotes all remaining flows.

As shown in Figure~\ref{fig:sens-samples}, accuracy rises sharply on all three datasets from 10 to 100 flows per class. CipherSpectrum-mix then saturates, whereas UNSW-TON-IoT and VisQUIC continue to improve up to 500 flows per class. With the full training set, accuracy decreases by 5.48 percentage points on UNSW-TON-IoT owing to class imbalance, changes only slightly on CipherSpectrum-mix, and increases by a further 4.99 percentage points on VisQUIC, where most packet content is protected, so the reported results on VisQUIC are conservative. Overall, a few hundred labeled flows per class suffice for most tasks, and the 500-flow bound, which follows ET-BERT~\cite{lin2022bert} and TrafficFormer~\cite{zhou2025trafficformer}, ensures a fair comparison with these baselines while avoiding the class imbalance of the full training set.

\section{Discussion}
\label{sec:discussion}
\paragraph{Implications.} As TLS~1.3 and QUIC continue to evolve and gain adoption, strong-encryption settings are likely to become increasingly relevant to practical traffic classification. The degradation of payload-driven methods observed in our experiments is consistent with the reduced availability of plaintext-derived information under modern encryption. The results suggest that the relative utility of protocol-semantic and behavioral information varies across traffic datasets, highlighting the importance of evaluation under diverse protocols, traffic types, and encryption settings.

\paragraph{Generalization.}The main experiments use random per-flow splits and therefore measure performance under a shared data distribution. We further examine FLARE under three types of distribution shift (Appendix~B). First, changing the cipher suite on CipherSpectrum introduces a substantial shift, reducing accuracy from 98.61\% to between 64.24\% and 77.46\%. Second, under simulated network perturbations, FLARE is robust to packet retransmission, with accuracy decreasing by at most 1.46 percentage points, whereas packet loss reduces accuracy by up to 9.56 percentage points at a 20\% loss rate. Third, on ISP backbone traffic from CESNET DataZoo, where only flow-level metadata is available and FLARE uses the directional packet-length representation alone, performance degrades gradually under temporal drift, with macro-F1 remaining at 75.16\% after nearly one year without retraining, and declines smoothly as the number of classes increases to more than one hundred.

\paragraph{Transfer to new networks.} FLARE can be transferred to a new network without re-pre-training. The frozen PacketEncoder is reused directly, and only the gated fusion module, FlowEncoder, and the classification head are fine-tuned on labeled flows from the target network, which typically requires a few hundred flows per class. If the target network carries protocols that the current parser does not support, the field-parsing rules need to be extended, and PacketEncoder may require supplementary pre-training on unlabeled traffic.

\paragraph{Limitations.} First, FLARE parses packets individually without TCP reassembly, so protocol fields beyond the retained bytes of each packet are not available to the semantic representation. Second, FLARE has not yet been evaluated on datasets that span multiple collection environments or provide long-term packet-level captures, and its generalization across environments and over time remains to be verified. Third, FLARE does not incorporate timing features, and its robustness to active obfuscation such as padding and traffic shaping has not been evaluated. Finally, the strongly encrypted datasets considered here focus on website fingerprinting and do not cover security tasks such as encrypted malware and intrusion detection.

\paragraph{Future Work.} Future work will investigate timing-aware behavioral representations, additional strong-encryption tasks, and lightweight packet encoders for high-throughput inference. We also plan to evaluate FLARE on packet-level captures spanning multiple environments and longer periods, and to develop field-level attribution, enabled by its field-aligned input, to clarify which observable information the model relies on. Finally, field-aware representations should be extended as emerging protocols continue to alter the set of observable header information.

\section{Conclusion}
\label{sec:conclusion}

The continued evolution of traffic encryption reduces the application content and high-level handshake information available to traffic classifiers, while the discriminative value of the remaining observable information varies across traffic scenarios. This paper introduces FLARE, which learns packet-level protocol-semantic representations through field-aware pre-training and adaptively integrates them with flow-level directional packet-length representations using a gating network. Experiments on diverse datasets show that FLARE consistently outperforms representative traditional machine-learning, deep-learning, and pre-trained baselines. Gating-weight analysis and ablation studies further indicate that FLARE captures scenario-dependent reliance on protocol-semantic and behavioral information. In strong-encryption settings involving TLS~1.3 and QUIC, FLARE achieves consistent improvements over both byte-driven and sequence-driven baselines. These findings confirm the value of adaptively integrating residual protocol semantics and observable traffic behavior for classification under constrained observability.

\section*{Acknowledgement}
This work was supported by National Natural Science Foundation of China (No. U25A6023) and the Cybersecurity Professional Committee of the Chinese Association of Educational Technology (NO. 2023CAET1002).

\bibliographystyle{unsrt}

\bibliography{cas-refs}

\clearpage

\appendix
\section{Field Parsing Statistics}
\label{app:parsing}

Parsing statistics are computed on the model input after applying the per-flow packet limit $N=64$ and the per-packet byte budget $B=80$, with packets parsed individually without TCP reassembly. Field-token coverage is the ratio of field tokens to the sum of field tokens and 2-byte fallback tokens. Encrypted TLS record bodies, encrypted QUIC payloads and protected header fields, and variable-length TCP options are retained as 2-byte tokens and therefore lower the coverage.

As shown in Table~\ref{tab:parsing-stats}, all capture files are read successfully, no packet raises a decoding error or yields malformed fields, and the IP and TCP/UDP header fields are parsed for every packet. QUIC public headers are parsed for all packets on VisQUIC; the unparsed QUIC packets on CIC-IoT-2022 are legacy gQUIC packets without explicit version identifiers, which are retained as 2-byte tokens rather than parsed with guessed field boundaries.

To evaluate TLS parsing independently of the parser, we reconstruct the TCP byte stream of each direction by sequence number and identify TLS record boundaries by requiring at least three consecutive valid records chained by their length fields. This reconstruction is used only for the audit. As shown in Table~\ref{tab:tls-audit}, 99.74\% of the record headers within the retained bytes are parsed in the downstream datasets, and only 6 of about 2.27 million parsed record headers are false positives. Record coverage is lower, at 50.56\%, because nearly all of the remaining record headers begin within a segment beyond the 80-byte budget, and thus fall outside the model input by design rather than being missed by the parser.

To examine parsing under real-world conditions, which may involve non-standard implementations, fragmentation, out-of-order delivery, and truncation, we apply the same procedure to traffic collected from a campus network, comprising 58,431 eligible flows from 102 application categories. As shown in the last rows of Tables~\ref{tab:parsing-stats} and~\ref{tab:tls-audit}, the IP and TCP/UDP header fields are again parsed for every packet, the overall field-token coverage is 62.67\%, and the parser recall of TLS record headers within the retained bytes is 99.62\% with no false positives.

\section{Generalization Evaluation}

\subsection{Cipher-Suite Shift and Network Perturbations}

\textbf{Cipher-suite shift.} We train FLARE on CipherSpectrum traffic encrypted with AES-128-GCM and test it on traffic encrypted with AES-256-GCM, ChaCha20-Poly1305 and on a mixed set, with the in-distribution AES-128-GCM test set as a reference. As shown in Table~\ref{tab:cross-suite}, accuracy decreases from 98.61\% to 64.24\%, 71.28\% and 77.46\%, respectively. A change of cipher suite therefore introduces a distribution shift, although accuracy remains well above the chance level.

\begin{table*}[htbp]
  \centering
  \caption{Field parsing statistics.}
  \label{tab:parsing-stats}
  \footnotesize
  \setlength{\tabcolsep}{3.5pt}
  \renewcommand{\arraystretch}{1.1}
  \begin{adjustbox}{max width=\linewidth,center}
    \begin{tabular}{@{}lrcccc@{}}
      \toprule
      \textbf{Dataset}
        & \textbf{Packets}
        & \makecell{\textbf{IP \& TCP/UDP}\\\textbf{headers parsed}}
        & \makecell{\textbf{QUIC headers}\\\textbf{parsed}}
        & \makecell{\textbf{TLS field-token}\\\textbf{coverage}}
        & \makecell{\textbf{Overall field-token}\\\textbf{coverage}} \\
      \midrule
      ISCX-NonVPN          & 13,835    & 100.00\% & --       & 14.74\% & 46.07\% \\
      CIC-IoT-2022-Attacks & 15,373    & 100.00\% & --       & --      & 47.90\% \\
      DoH-Malicious        & 1,636,486 & 100.00\% & --       & 21.26\% & 63.29\% \\
      \midrule
      CSTNET-TLS~1.3       & 212,028   & 100.00\% & --       & 9.34\%  & 51.63\% \\
      ISCX-VPN-Service     & 78,586    & 100.00\% & --       & 13.69\% & 48.52\% \\
      ISCX-VPN-App         & 72,371    & 100.00\% & --       & 12.86\% & 48.26\% \\
      UNSW-TON-IoT         & 72,066    & 100.00\% & --       & 16.06\% & 46.34\% \\
      USTC-TFC             & 2,327,218 & 100.00\% & --       & 6.48\%  & 46.88\% \\
      CIC-IoT-2022         & 48,208    & 100.00\% & 34.18\%  & 14.77\% & 49.33\% \\
      CipherSpectrum-mix   & 1,023,452 & 100.00\% & --       & 9.79\%  & 52.52\% \\
      VisQUIC              & 482,922   & 100.00\% & 100.00\% & --      & 37.84\% \\
      \midrule
       Dataset total     & 7,340,578 & 100.00\% & 99.74\%  & 9.83\%  & 49.65\% \\
      Campus network       & 1,257,844 & 100.00\% & --       & 16.26\% & 62.67\% \\
      \bottomrule
    \end{tabular}
  \end{adjustbox}
\end{table*}

\begin{table}[htbp]
  \centering
  \caption{Independent audit of TLS record-header parsing.}
  \label{tab:tls-audit}
  \footnotesize
  \setlength{\tabcolsep}{4pt}
  \renewcommand{\arraystretch}{1.1}
  \begin{adjustbox}{max width=\linewidth,center}
    \begin{tabular}{@{}lcccc@{}}
      \toprule
      \textbf{Dataset}
        & \makecell{\textbf{Parser}\\\textbf{recall}}
        & \makecell{\textbf{False}\\\textbf{positives}}
        & \makecell{\textbf{Record}\\\textbf{coverage}}
        & \makecell{\textbf{Payload-start}\\\textbf{visibility}} \\
      \midrule
      ISCX-NonVPN        & 99.94\%  & 0.00\%   & 75.95\% & 73.07\% \\
      DoH-Malicious      & 100.00\% & 0.00\%   & 88.01\% & 94.99\% \\
      \midrule
      CSTNET-TLS~1.3     & 99.03\%  & 0.00\%   & 64.41\% & 50.94\% \\
      ISCX-VPN-Service   & 99.98\%  & 0.02\%   & 77.39\% & 61.58\% \\
      ISCX-VPN-App       & 99.98\%  & 0.02\%   & 77.55\% & 58.52\% \\
      UNSW-TON-IoT       & 99.71\%  & 0.00\%   & 65.40\% & 65.31\% \\
      USTC-TFC           & 99.30\%  & 0.00\%   & 52.73\% & 26.93\% \\
      CIC-IoT-2022       & 99.73\%  & 0.00\%   & 75.53\% & 60.23\% \\
      CipherSpectrum-mix & 99.76\%  & 0.00\%   & 49.50\% & 57.69\% \\
      \midrule
      Dataset total   & 99.74\%  & 0.0003\% & 50.56\% & 57.06\% \\
      Campus network     & 99.62\%  & 0.00\%   & 60.37\% & 59.51\% \\
      \bottomrule
    \end{tabular}
  \end{adjustbox}

  \vspace{2pt}
  \begin{minipage}{\linewidth}
    \scriptsize
    Parser recall is the proportion of record headers within the retained bytes that are parsed; record coverage is the proportion of all record headers carried by the retained packets that are parsed; payload-start visibility is the proportion of payload-bearing TLS packets whose payload begins with a record header.
  \end{minipage}
\end{table}

\begin{table}
    \centering
    \small
    \caption{Cross-cipher-suite evaluation on CipherSpectrum.}
    \begin{tabular}{lccc}
    \toprule
    Test Setting & Test suite & Acc & F1 \\
        \midrule
        In-distribution & AES-128-GCM & 0.9861 & 0.9862\\
        Cross-suite & AES-256-GCM & 0.6424 & 0.6540\\
        Cross-suite & ChaCha20-Poly1305 & 0.7128 & 0.6996\\
        Cross-suite & Mix & 0.7746 & 0.7760\\
    \bottomrule
    \end{tabular}
    \label{tab:cross-suite}
\end{table}

\begin{figure}
    \centering

    \begin{subfigure}[t]{0.48\columnwidth}
        \centering
        \includegraphics[width=\linewidth]
            {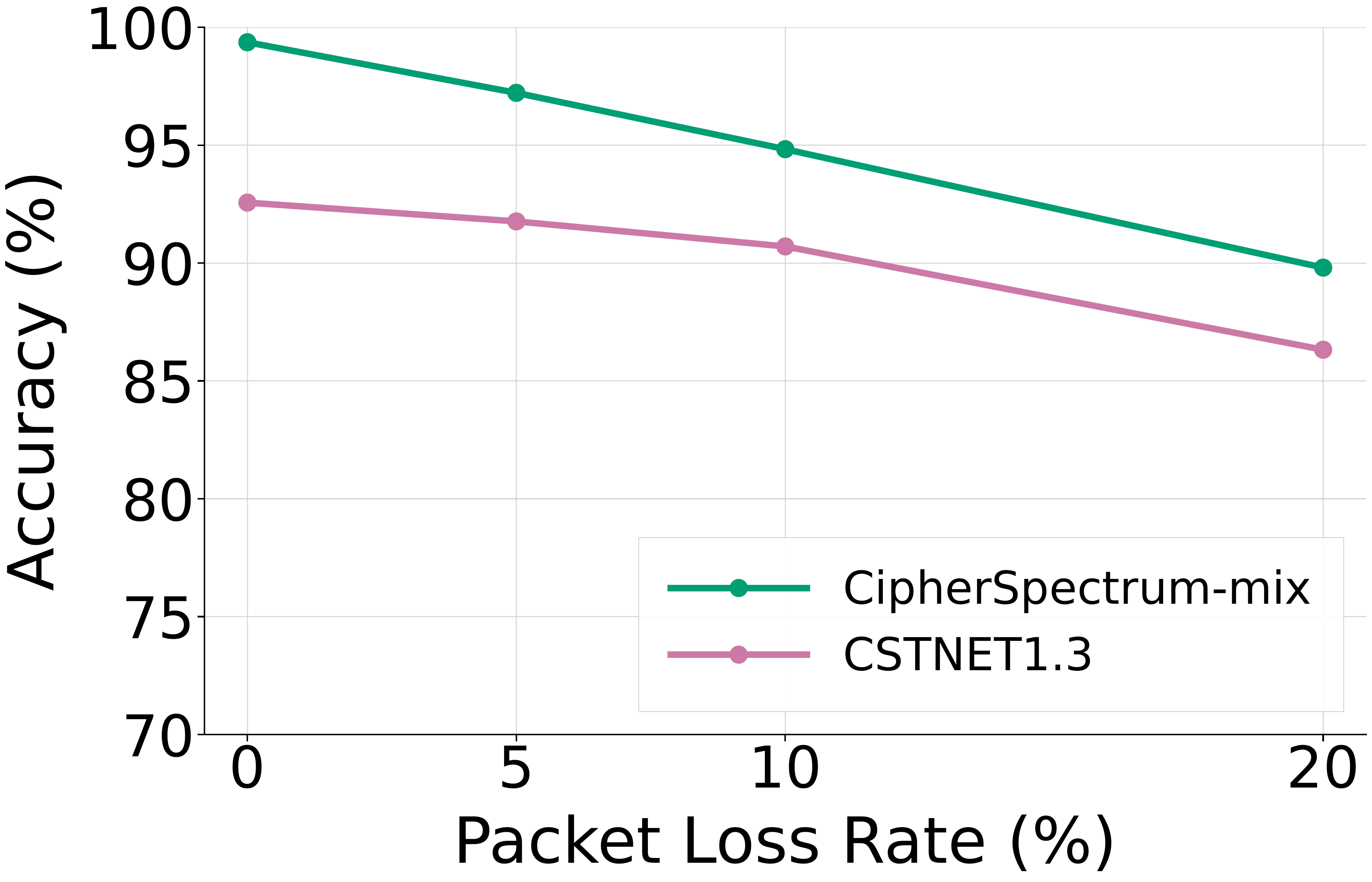}
        \caption{Packet loss}
        \label{fig:loss}
    \end{subfigure}
    \hfill
    \begin{subfigure}[t]{0.48\columnwidth}
        \centering
        \includegraphics[width=\linewidth]
            {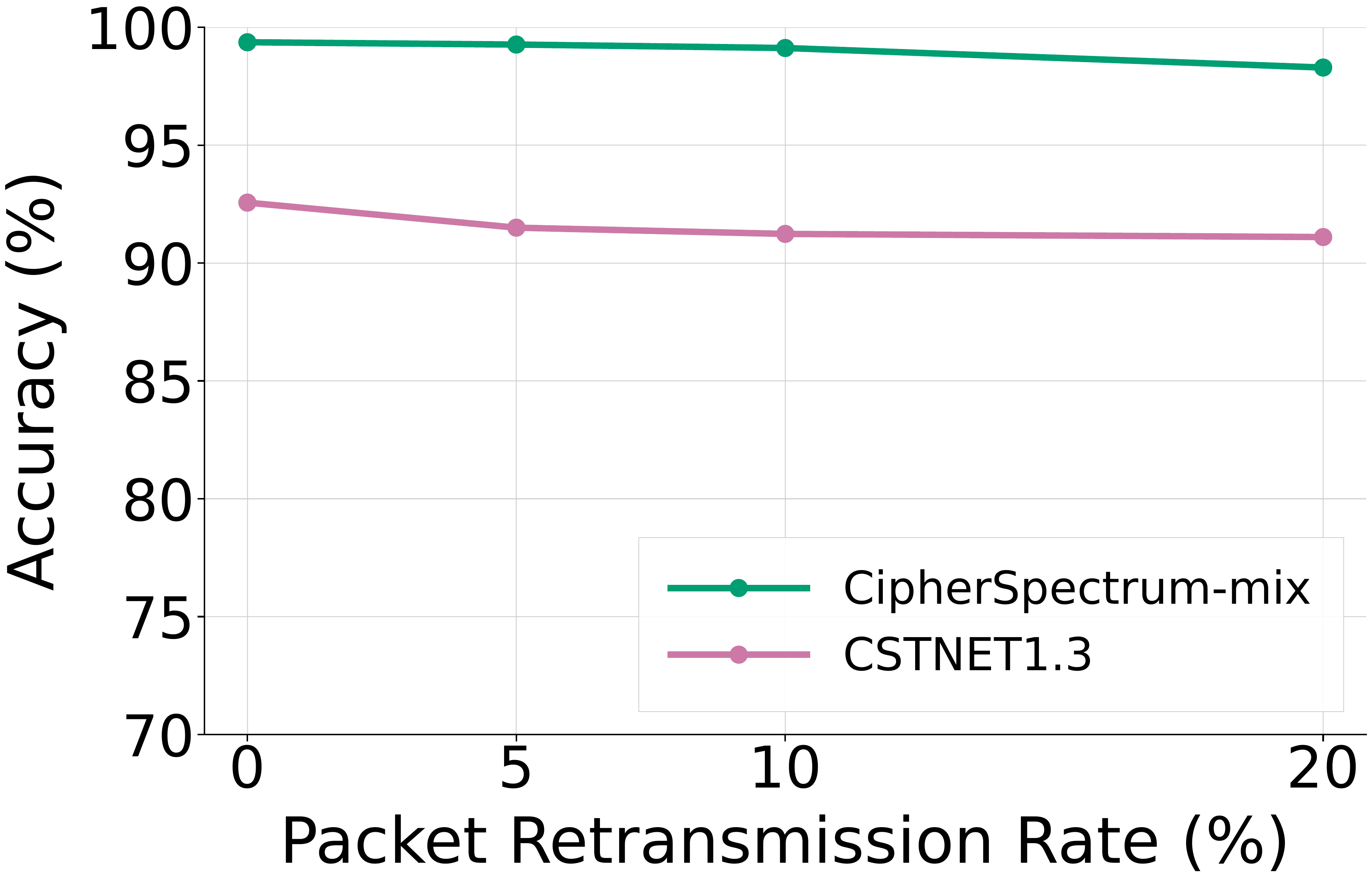}
        \caption{Packet retransmission}
        \label{fig:retransmission}
    \end{subfigure}

    \caption{Accuracy of FLARE under test-time network perturbations on CipherSpectrum-mix and CSTNET-TLS~1.3. (a) Each packet is dropped independently with probability~$p$. (b) Each packet is followed by an identical copy with probability~$p$.}
    \label{fig:perturbation}
\end{figure}

\textbf{Network perturbations.} To approximate changes in network conditions, we perturb only the test flows of CipherSpectrum-mix and CSTNET-TLS~1.3, leaving the trained model unchanged. For packet loss, each packet is dropped independently with probability $p$; for retransmission, each packet is followed by an identical copy with probability $p$. In both cases, the first 64 packets of the perturbed flow are used as input, and $p \in \{5\%, 10\%, 20\%\}$. As shown in Figure~\ref{fig:perturbation}, FLARE is robust to retransmission, with accuracy decreasing by at most 1.46 percentage points even at $p = 20\%$, since duplicated packets preserve the original packets and their order. Packet loss has a larger effect, as it removes packets and shifts the positions of subsequent ones. Accuracy decreases by up to 9.56 and 6.24 percentage points on the two datasets at a loss rate of 20\%, but by only 2.15 and 0.80 percentage points at 5\%.

\subsection{Metadata-Only ISP Traffic}
The datasets in the main experiments provide raw packets but cover limited periods and moderate numbers of classes. Large-scale deployments such as ISP backbone networks must additionally cope with long-term traffic evolution and much larger service sets, often with only flow-level metadata retained. We therefore evaluate FLARE on the CESNET DataZoo datasets, captured on an ISP backbone network over periods of up to one year. Since only flow-level metadata is available, PacketEncoder cannot be applied, and FLARE uses the directional packet-length representation alone. These evaluations thus examine the sequential representation of FLARE under temporal drift and class-space scaling.

\subsubsection{Temporal Drift}
Following the MFWDD study~\cite{janvcivcka2024mfwdd}, we use CESNET-QUIC22~\cite{luxemburk2023cesnet} and CESNET-TLS-Year22~\cite{hynek2024cesnet} for the temporal drift evaluation. We selected 10 service classes, sampled 500 flows per class, and split the data into training and test sets by time.

Table~\ref{tab:short-term drift} reports the short-term drift results. On CESNET-QUIC22, macro-F1 decreases by a cumulative 10.14 percentage points over three weeks, whereas CESNET-TLS-Year22 shows no noticeable decline over the same period.

\begin{table*}
    \centering
    \small
    \caption{Short-term drift evaluation on CESNET-QUIC22 and CESNET-TLS-Year22. Values are macro-F1 (\%). Changes relative to the in-period result are given in parentheses (pp).}
    \begin{tabular}{lcccccc}
    \toprule
    Dataset & Classes & Train week & $+0$ week & $+1$ week & $+2$ weeks & $+3$ weeks \\
        \midrule
        CESNET-QUIC22     & 10 & W-2022-44 & 85.27 & 80.96 ($-$4.31) & 75.34 ($-$9.93) & 75.13 ($-$10.14) \\
        CESNET-TLS-Year22 & 10 & W-2022-02 & 95.40 & 95.96 ($+$0.56) & 95.77 ($+$0.37) & 96.28 ($+$0.88) \\
    \bottomrule
    \end{tabular}
    \label{tab:short-term drift}
\end{table*}

\begin{figure}
    \centering
    \includegraphics[width=1\columnwidth]{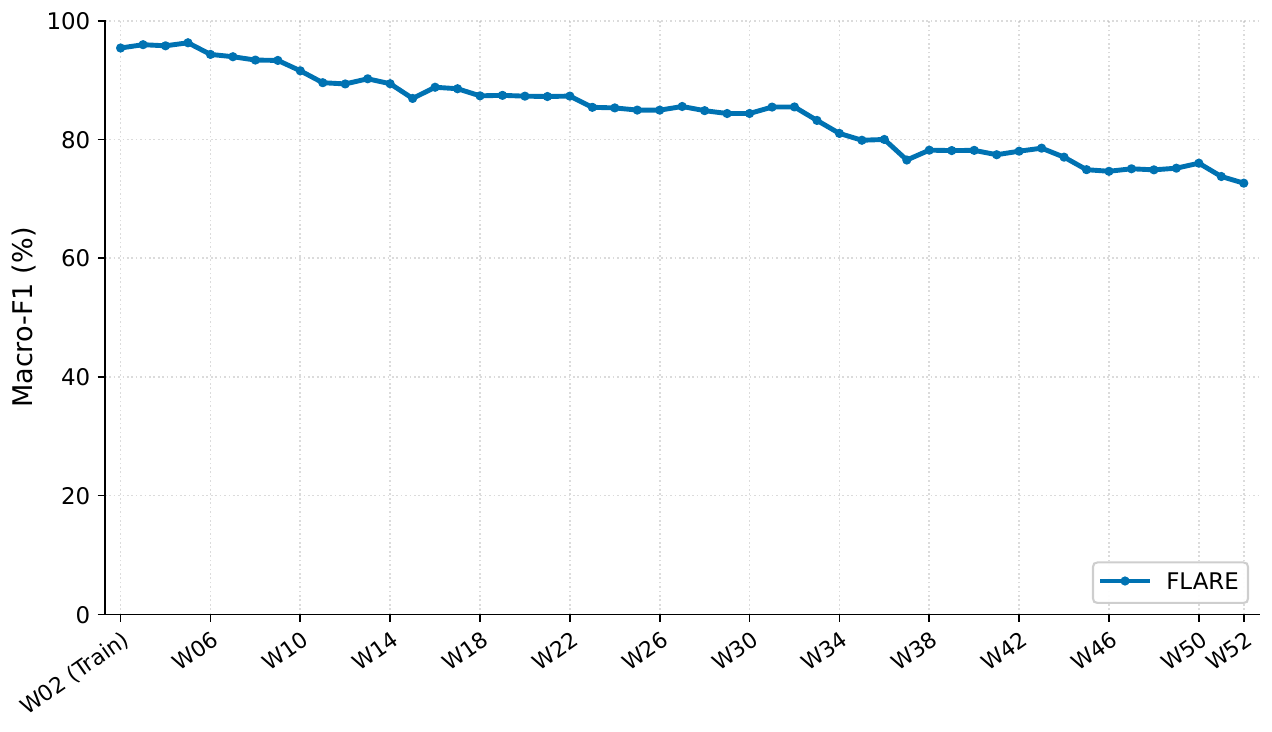}
    \caption{Weekly macro-F1 of FLARE under temporal drift on CESNET-TLS-Year22.}
    \label{fig:long-term drift}
\end{figure}

The long-term drift results are reported in Table~\ref{tab:long-term drift} and Figure~\ref{fig:long-term drift}. When trained on week W-2022-02 of CESNET-TLS-Year22, FLARE degrades gradually and still achieves a macro-F1 of 75.16\% after nearly one year.

\begin{table}
    \centering
    \small
    \caption{Long-term drift evaluation on CESNET-TLS-Year22.  $\Delta$ denotes the change in macro-F1 relative to the in-period result.}
    \begin{tabular}{ccccc}
    \toprule
        Week & Gap (weeks) & Acc & F1 & $\Delta$ F1 (pp) \\
        \midrule
        W-2022-02 & 0  & 0.9540 & 0.9540 & 0.00 \\
        W-2022-18 & 16 & 0.8749 & 0.8735 & $-$8.06 \\
        W-2022-35 & 33 & 0.8000 & 0.7986 & $-$15.54 \\
        W-2022-49 & 47 & 0.7561 & 0.7516 & $-$20.24 \\
        \bottomrule
    \end{tabular}
    \label{tab:long-term drift}
\end{table}

\subsubsection{Class-Space Scaling}
To examine how FLARE performs on a large public dataset, we conducted class-scaling experiments on four CESNET DataZoo datasets: CESNET-QUIC22~\cite{luxemburk2023cesnet}, CESNET-TLS-Year22~\cite{hynek2024cesnet}, CESNET-TLS22~\cite{luxemburk2023fine}, and CESNET-QUICEXT-25~\cite{luxemburk2025cesnet}. Each dataset was fixed to a single collection period, namely W-2022-44, W-2022-02, W-2021-40, and M-2025-1, respectively. As in the temporal drift evaluation, FLARE uses the directional packet-length pathway alone, taking the signed lengths of the first 30 packets of each flow as input. To exclude the effect of class imbalance, we sampled 1,000 flows per class and split them 8:1:1 into training, validation, and test sets. Class subsets were taken as prefixes of a single random permutation, so that each smaller subset is contained in every larger one, and the number of classes was increased from 10 to the full class set of each dataset.

\begin{figure}
    \centering
    \includegraphics[width=1\columnwidth]{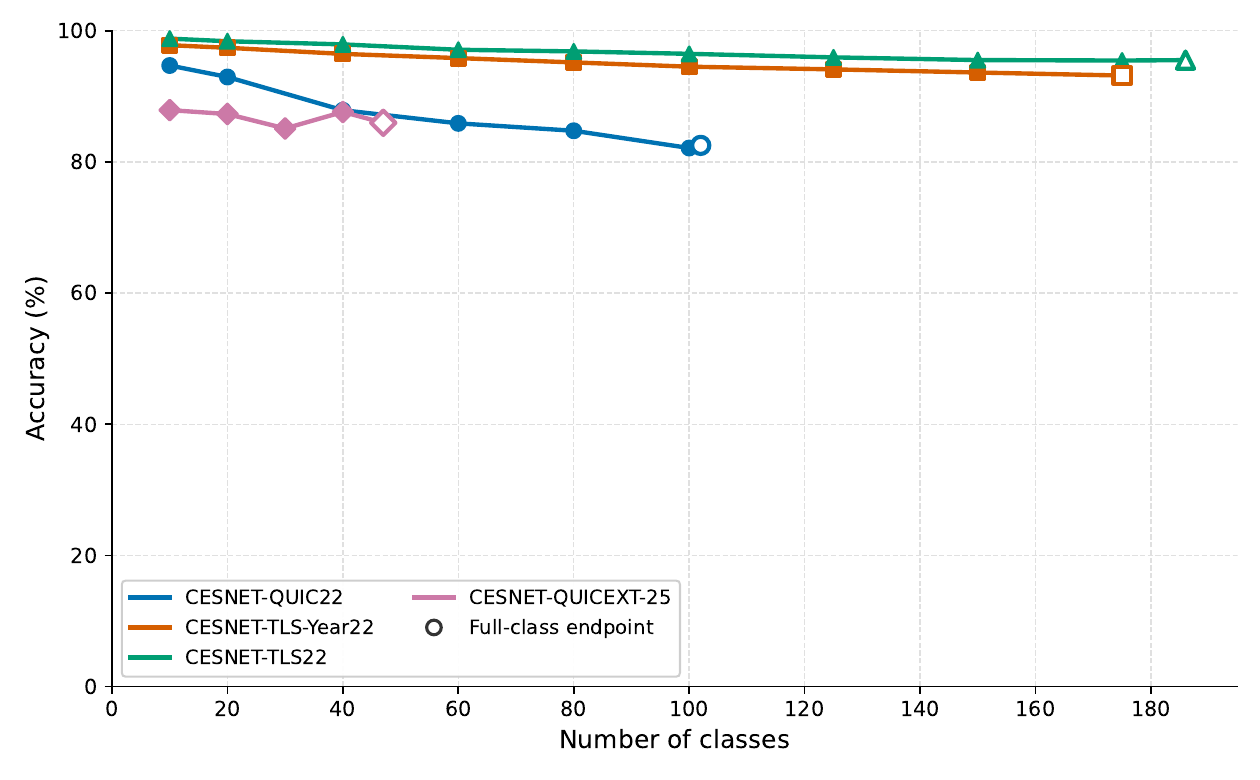}
    \caption{Accuracy of FLARE with an increasing number of classes on four CESNET datasets.}
    \label{fig:class_count_scaling}
\end{figure}

    The results are shown in Figure~\ref{fig:class_count_scaling}. As the number of classes increases, the accuracy of FLARE declines steadily without abrupt drops. When expanding from 10 classes to the full class set, accuracy decreases by 4.63 percentage points on CESNET-TLS-Year22 (175 classes), 3.28 percentage points on CESNET-TLS22 (186 classes), and 1.94 percentage points on CESNET-QUICEXT-25 (47 classes). CESNET-QUIC22 is more sensitive to class-space growth, with accuracy falling from 94.70\% to 82.52\%, a drop of 12.18 percentage points. The slight increases between some adjacent settings, such as from 30 to 40 classes on CESNET-QUICEXT-25 and from 100 to 102 classes on CESNET-QUIC22, arise from variation across independently trained models and do not alter the overall downward trend. These results indicate that directional packet-length sequences remain discriminative on real ISP traffic with more than one hundred service classes.

\end{document}